\documentclass{jfm}
\usepackage{graphicx}
\usepackage{epstopdf, epsfig}
\usepackage{amsmath}
\usepackage{siunitx}
\usepackage{pifont}
\usepackage{dcolumn}
\usepackage{makecell}

\shorttitle{Intricate Role of Thermal Properties and Volatility in Droplet Spreading}
\shortauthor{Z. Wang, G. Karapetsas, P. Valluri and C. Inoue}

\title{Intricate Role of Thermal Properties and Volatility in Droplet Spreading: A Generalization to Tanner’s Law}

\author{Zhenying Wang\aff{1}\aff{2}
  \corresp{\email{zhenying.wang@aero.kyushu-u.ac.jp}},
  George Karapetsas\aff{3}
  Prashant Valluri\aff{4}
  \and Chihiro Inoue\aff{1}}

\affiliation{\aff{1}Department of Aeronautics and Astronautics, Kyushu University,
Nishi-Ku, Motooka 744, Fukuoka 819-0395, Japan
\aff{2}International Institute for Carbon-Neutral Energy Research (WPI-I$^2$CNER), Kyushu University, Nishi-Ku, Motooka 744, Fukuoka 819-0395, Japan
\aff{3}Department of Chemical Engineering, Aristotle University of Thessaloniki, Thessaloniki 54124, Greece
\aff{4}Institute of Multiscale Thermofluids, School of Engineering, University of Edinburgh, Edinburgh EH9 3JL, UK}

\begin{document}

\maketitle

\begin{abstract}
Droplet spreading is ubiquitous and plays a significant role in liquid-based energy systems, thermal management devices, and microfluidics. While the spreading of non-volatile droplets is quantitatively understood, the spreading and flow transition in volatile droplets remains elusive due to the complexity added by interfacial phase change and non-equilibrium thermal transport. Here we show, using both mathematical modeling and experiments, that the wetting dynamics of volatile droplets can be scaled by the spatial-temporal interplay between capillary, evaporation, and thermal Marangoni effects. We elucidate and quantify these complex interactions using phase diagrams based on systematic theoretical and experimental investigations. A spreading law of evaporative droplets is derived by generalizing Tanner’s law (valid for non-volatile liquids) to a full range of liquids with saturation vapor pressure spanning from $10^1$ to $10^4$ Pa and on substrates with thermal conductivity from $10^{-1}$ to $10^3$ W/m/K. Besides its importance in fluid-based industries, the conclusions also enable a unifying explanation to a series of individual works including the criterion of flow reversal and the state of dynamic wetting, making it possible to control liquid transport in diverse application scenarios. 
\end{abstract}

\begin{keywords}
Authors should not enter keywords on the manuscript, as these must be chosen by the author during the online submission process and will then be added during the typesetting process (see http://journals.cambridge.org/data/\linebreak[3]relatedlink/jfm-\linebreak[3]keywords.pdf for the full list)
\end{keywords}

\section{Introduction}
Wetting presents the basic state when a finite volume of liquid contacts a solid substrate (\cite{de1985wetting}, \cite{bonn2009wetting}). The requirement for wetting differs for different applications, and hence the design principles. For example, good wetting enables efficient liquid supply and avoids dry-out in phase change thermal management, \emph{e.g.}, electronics cooling (\cite{mudawar2001assessment}). While in condensation, a quick wetting transition will increase the substrate’s thermal resistance and worsen the condensation performance (\cite{rose2002dropwise}). A full understanding of the interplay between wetting and phase change is thus necessary for in-demand design of various energy systems (\cite{tu2018progress}, \cite{breitenbach2018drop}) as well as in other microfluids based industries (\cite{zheng2010directional}, \cite{teh2008droplet}). 

The wetting dynamics of nonvolatile liquids are known to be  governed by capillary force and viscous dissipation. Specifically, the spreading rate of a nonvolatile Newtonian droplet in perfect wetting cases can be quantitatively described by Tanner’s law, $R(t)\sim t^{1/10}$ (\cite{tanner1979spreading}, \cite{lelah1981spreading}, \cite{blake2006physics}), which can be derived mathematically by employing the lubrication approximation and balancing the effect of capillary forces with the resisting viscous forces generated by the droplet motion (\cite{carlson2018fluctuation}, \cite{pang2022mathematical}, \cite{voinov1976hydrodynamics}). The spreading law was then validated by successive experiments (\cite{marmur1983equilibrium}) and further extended to non-Newtonian liquids (\cite{rafai2004spreading}, \cite{jalaal2021spreading}), gravity-influenced cases (\cite{cazabat1986dynamics}), near-critical cases (\cite{saiseau2022near}), liquids on non-rigid solid and on thin liquid films (\cite{carre1996viscoelastic}, \cite{cormier2012beyond}), as well as under-liquid systems (\cite{mitra2016understanding}, \cite{goossens2011can}). 

Compared to nonvolatile liquids, the spreading of volatile liquids is more complex and remains elusive owing to the complexity added by interfacial phase change and non-equilibrium thermal transport (\cite{jambon2018spreading}, \cite{wang2022wetting}, \cite{berteloot2008evaporation}, \cite{wang2019coupled}). Specifically, the non-uniform mass flux tends to recede the liquid-air interface especially near the three phase contact line. The effect of evaporation cooling and the non-uniform pathway for heat dissipation induces temperature gradient and therefore thermal Marangoni flow, which complexes the flow pattern inside the droplet and further affects the spreading dynamics. 

A full elucidation of these interacting mechanisms requires a quantitative description of each physical process, as well as a thorough understanding of all influencing parameters. In this research, we combine experimental and numerical approaches to address these issues and revisit the milestone descriptions of flow interaction in liquid spreading. The explorations enable us to quantify the strength of dominating physics that varies spatially and temporally, which on the other hand provides a unifying criterion for flow reversal (\cite{hu2005analysis}, \cite{xu2007marangoni}, \cite{ristenpart2007influence}, \cite{xu2010criterion}), pattern formation (\cite{hu2006marangoni}, \cite{deegan1997capillary}, \cite{li2015coffee}), and Marangoni-capillary interactions (\cite{tsoumpas2015effect}, \cite{shiri2021thermal}, \cite{yang2022evaporation}) in drying droplets. 

Specifically, we investigate the flow state and the resulting wetting dynamics of volatile liquids on thermal conductive substrates. In the numerical aspect, previous lubrication-type models usually consider boundary conditions with uniform heating (\cite{anderson1995spreading}, \cite{ajaev2005spreading}) or preset substrate temperature/temperature gradient (\cite{karapetsas2013effect}, \cite{dai2016thermocapillary}, \cite{charitatos2020thin}, \cite{wang2021dynamics}), which failed to illustrate the non-negligible role of non-isothermal heat conduction in the overall process. Additionally, the numerical and experimental investigations have at large been developing in parallel without strict comparisons that consider the specific application scenarios.  In this work we take account of the mass, momentum, and energy transport in the liquid phase, as well as the heat conduction through the solid substrate. We apply the assumption of a precursor film (ultrathin adsorbed liquid layer, \cite{kavehpour2003microscopic}, \cite{hoang2011dynamics}, \cite{ajaev2005spreading}) existing in front of the three-phase contact line, which eliminates the stress singularity and avoids experimental inputs on the contact line motion, \emph{e.g.}, preset advancing/receding contact angles or a finite slip. A strict one-to-one comparison is then conducted between experimental and numerical results for a broad range of liquid-solid combinations. The reliability of the theoretical models further give rise to a general principle of flow transition and non-isothermal spreading of volatile droplets.

The rest of this paper is organized as follows. In \S \ref{sec:model}, we introduce the formulated mathematical model, the scaling and simplification, and the numerical method. \S \ref{sec:experimental_method} introduces the material preparation, the detailed experimental setup and procedures. \S \ref{sec:results_discussion} presents the numerical and experimental results including the dominating mechanisms, the transition of flow pattern, and their relation to the spreading dynamics. \S \ref{sec:conclusions} summarizes the key findings of this work, and share our perspectives on the spreading of Newtonian/non-Newtonian fluids with/without interfacial phase change.

\section{Formulation of Numerical Model}\label{sec:model}
We consider a thin liquid droplet (aspect ratio $\varepsilon = \hat{H}_0/\hat{R}_0\ll1$, incompressible Newtonian fluid) sitting on a thermal conductive solid substrate with finite thickness and surrounded by a mixture of its own vapor and noncondensing gas. The low aspect ratio allows for the application of lubrication theory which simplifies the Navier-Stokes equations while effectively describes the dominating physical processes. The lubrication approximation is applicable for droplets with small and moderate contact angles typically less than 40°, which is suitable for current research on complete wetting and partial wetting cases. A precursor film is assumed to exist at the solid surface in front of the three phase contact line. This avoids the stress singularity that may arise due to the inherent contradiction in simultaneously assuming the no-slip boundary condition and expecting a displacement between liquid and gas at the moving contact line. 
\begin{figure}
\centering
\includegraphics[width=13cm]{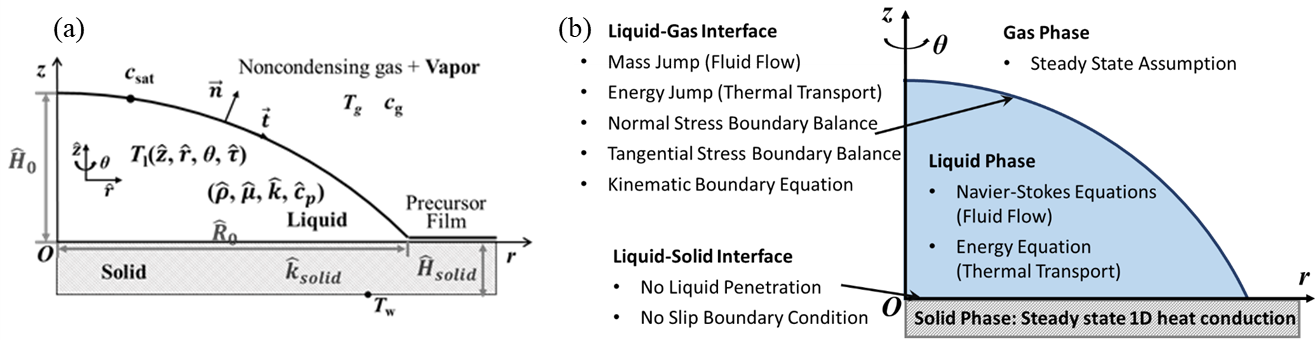}
\caption{(1) A single component droplet sitting on a solid substrate in contact with its own vapor and noncondensing gas: The aspect ratio, $\hat{H}_0/\hat{R}_0$, is assumed as very small, $c_\mathrm{sat}$ denotes the saturation vapor concentration at the liquid-air interface, $c_\mathrm{g}$ refers to the vapor concentration in the gas phase and keeps constant, $\hat{H}_\mathrm{solid}$ denotes the substrate thickness and $\hat{k}_\mathrm{solid}$ denotes the thermal conductivity. $\vec n$ and $\vec t$ denote the outward units vectors acting in normal and tangential directions to the interface. The center of droplet base, \emph{O}, is defined as the origin of the coordinate. (2) Governing equations in the gas, liquid and solid phases along with the boundary equations at the central axis, the liquid-gas interface and the liquid-solid interface.}
\label{fig:model}
\end{figure}

\textbf{Governing Equations} A cylindrical coordinate system, $(\hat{r},\theta,\hat{z})$, is applied to solve the velocity field, $\hat{u}=(\hat{u},\hat{v},\hat{w})$, where $\hat{u}$, $\hat{v}$, and $\hat{w}$ correspond to the horizontal, azimuthal and vertical components of the velocity field respectively. The liquid-vapor interface locates at $\hat{z}=\hat{h}(\hat{r},\hat{t})$ and the liquid-solid and solid-vapor interfaces locate at $\hat{z}=0$. The liquid phase is governed by the incompressible mass, momentum and energy equations, Eqs. (\ref{eq:mass}, \ref{eq:momentum}, \ref{eq:energy}),
\begin{equation}
    \hat{\nabla}\cdot \hat{\boldsymbol{u}} = 0,
    \label{eq:mass}
\end{equation}
\begin{equation}
    \hat{\rho}(\frac{\partial\hat{\boldsymbol{u}}}{\partial\hat{t}}+\hat{\boldsymbol{u}}\cdot \hat{\nabla} \hat{\boldsymbol{u}})=\hat{\nabla} \cdot \hat{\boldsymbol{E}},
    \label{eq:momentum}
\end{equation}
\begin{equation}
    \hat{\rho}\hat{c}_p(\frac{\partial\hat{T}}{\partial\hat{t}}+\hat{\boldsymbol{u}}\cdot \hat{\nabla} \hat{T})=\hat{\nabla} \cdot \hat{k}\hat{\nabla}\hat{T},
    \label{eq:energy}
\end{equation}
where liquid density $\hat{\rho}$, specific heat capacity $\hat{c}_p$, and thermal conducitivity $\hat{k}$ are set as constant, $\hat{T}$ denotes temperature, $\hat{t}$ denotes time, and $\hat{\boldsymbol{E}}$ denotes the total stress tensor at the liquid side,
\begin{equation}
    \hat{\boldsymbol{E}}=-\hat{p}\boldsymbol{I}+\hat{\mu}(\hat{\nabla}\hat{\boldsymbol{u}}+\hat{\nabla}\hat{\boldsymbol{u}}^T),
    \label{eq:stress_tensor}
\end{equation}
where $\hat{\mu}$ is the dynamic viscosity (linear function of liquid temperature), and $\boldsymbol{I}$ is the identity tensor.

At the liquid-gas interface, the liquid velocity and geometry satisfy a force balance both in the normal and tangential directions. The normal stress balance relates the normal stress jump, surface tension, mean curvature and van der Waals interactions,
\begin{equation}
    -\hat{p}+\boldsymbol{n}\cdot \hat{\boldsymbol{\tau}}\cdot \boldsymbol{n}=2\hat{\kappa}\hat{\sigma}+\hat{\Pi}-\hat{p}_g,
    \label{eq:normal_stress_balance}
\end{equation}
where $\hat{p}$ is the pressure of the liquid phase, $\hat{p}_g$ is the total pressure of the gas phase, $\hat{\boldsymbol{\tau}}$ is the shear stress tensor of the liquid phase. $2\hat{\kappa}=-\hat{\nabla}_S\cdot\boldsymbol{n}$ is twice the mean curvature of the free surface and $\hat{\nabla}_S=(\boldsymbol{I}-\boldsymbol{n}\boldsymbol{n})\cdot\hat{\nabla}$ is the surface gradient operator. $\hat{\Pi}$ denotes the disjoining pressure accounting for intermolecular interactions near the contact line (\cite{de1990dynamics}),
\begin{equation}
    \hat{\Pi}=\frac{\hat{A}}{6\pi\hat{h}^3},
    \label{eq:disjoining_pressure}
\end{equation}
with $\hat{A}$ being the dimensional Hamaker constant.

The tangential stress balance relates the shear stress jump and the surface tension gradient. By ignoring the shear stress from the gas phase due to small viscosity, we arrive at,
\begin{equation}
\boldsymbol{n}\cdot\hat{\boldsymbol{E}}\cdot\boldsymbol{t}=\hat{\nabla}_S\hat{\sigma}\cdot\boldsymbol{t},
\label{eq:tangential_stress_balance}
\end{equation}
where $\boldsymbol{t}$ refers to the outward vectors which is tangential to the interface.

The motion of free surface can be described with the kinematic boundary condition, expressed as,
\begin{equation}
\frac{\partial\hat{h}}{\partial\hat{t}}+\hat{u}_S\frac{\partial\hat{h}}{\partial\hat{r}}+\frac{\hat{v}_S}{\hat{r}}\frac{\partial\hat{h}}{\partial\theta}=\hat{w}_S.
\label{eq:kinematic_boundary}
\end{equation}

Along the droplet interface, the boundary equation of mass balance can be expressed by the relationship between the velocity of the liquid solution, $\boldsymbol{\hat{u}}$, and the velocity of the interface, $\boldsymbol{\hat{u}}_s$,
\begin{equation}
(\boldsymbol{\hat{u}}-\boldsymbol{\hat{u}}_S)\cdot n=\frac{\hat{J}}{\hat{\rho}},
\label{eq:mass_balance_boundary}
\end{equation}
where $\hat{J}$ denotes the interfacial mass flux of water vapor. The jump energy balance can then be derived taking account of the heat release at the liquid-vapor interface and ignoring the heat conduction into the gas phase,
\begin{equation}
    \hat{J}\hat{L}+\hat{k}\hat{\nabla}\hat{T}\cdot\boldsymbol{n}=0,
    \label{eq:jump_energy_balance}
\end{equation}
where $\hat{L}$ denotes the latent heat of vaporization.
The Hertz–Knudsen equation is commonly used for predicting the mass flux induced by evaporation or condensation towards a liquid–vapor interface. At thermodynamic equilibrium, the chemical potentials of the gas phase and liquid phase across the liquid–air interface reach balance, and the following relation can be derived combining the Hertz–Knudsen equation (\cite{plesset1976flow}, \cite{moosman1980evaporating}) and the chemical balance relations (\cite{atkins2014atkins}).
\begin{equation}
    \hat{J}=\hat{p}_\mathrm{v,sat}\sqrt{\frac{\hat{M}}{2\pi\hat{R}_g\hat{T}_g}}(\frac{\hat{M}}{\hat{\rho}\hat{R}_g\hat{T}_g}(\hat{p}-\hat{p}_g)+\frac{\hat{M}\hat{L}}{\hat{R}_g\hat{T}_g^2}(\hat{T}_i-\hat{T}_g)+ln(\frac{1}{\chi_\mathrm{vapor}})),
    \label{eq:mass_flux}
\end{equation}
where $\hat{M}$ is the molar mass of water, $\hat{R}_g$ is the gas constant, $\hat{T}_g$ is the temperature of gas phase, $\hat{T}_i$ is the temperature at the liquid-gas interface, $\hat{p}_\mathrm{v,sat}$ is the saturation vapor pressure in the gas phase, and   is the relative vapor concentration, defined as the ratio of vapor pressure to the saturation vapor pressure in the gas phase.

\textbf{Scaling} The key parameters are scaled as follows,
\begin{equation}
\begin{split}
\hat{r}=\hat{R}_0 r,\ \hat{z}=\hat{H}_0 z,\ (\hat{u},\hat{w})=(\hat{u}^*u,\frac{\hat{H}_0}{\hat{R}_0}\hat{u}^*w),\\ \hat{\mu}=\hat{\mu}_0\hat{\mu},\ \hat{T}=\hat{T}_\mathrm{ref}+T\Delta \hat{T}, \
\hat{\sigma}=\hat{\sigma}_0\sigma=\hat{\sigma}_0(1+\frac{\hat{\eta}_\sigma \Delta\hat{T}}{\hat{\sigma}_0}T),\\
\hat{c}_p=\hat{c}_{p,0}c_p, \ \hat{k}=\hat{k}_0 k,\ \hat{t}=\frac{\hat{R}_0}{\hat{u}^*}t,\ \hat{p}=\hat{p}_g+\frac{\hat{\mu}_0\hat{u}^*\hat{R}_0}{\hat{H}_0^2}p,\ \hat{J}=\frac{\hat{k}_0\Delta\hat{T}}{\hat{H}_0\hat{L}}J.
\end{split}
\end{equation}
We define characteristic velocity as $\hat{u}^*=\frac{\varepsilon\hat{\eta}_\sigma \Delta\hat{T}}{\hat{\mu}_0}$  and derive the dimensionless group, including the Reynolds number, $\mathrm{Re}=\frac{\hat{\rho}_0\hat{u}^*\hat{H}_0}{\hat{\mu}_0}$, Stokes number, $\mathrm{St}=\frac{\hat{\rho}_0\hat{g}_G\hat{H}_0^3}{\hat{\mu}_0\hat{u}^*\hat{R}_0}$, Prandtl number, $\mathrm{Pr}=\frac{\hat{\mu}_0\hat{c}_{p,0}}{\hat{k}_0}$, evaporation number, $E=\frac{\hat{k}_0\hat{R}_0\Delta\hat{T}}{\hat{\rho}_0\hat{L}\hat{H}_0^2\hat{u}^*}$, Marangoni number, $\mathrm{Ma}=\frac{\hat{\mu}_0\hat{u}^*}{\varepsilon\hat{\sigma}_0}=\frac{\hat{\eta}_\sigma\Delta\hat{T}}{\hat{\sigma}_0}$, Hamaker constant, $\mathcal{A}=\frac{\hat{\mathcal{A}}}{6\pi\hat{\mu}_0\hat{u}^*\hat{R}_0\hat{H}_0}$, the coefficients in the expression of mass flux, $\delta=\frac{\hat{M}\hat{\eta}_\sigma\Delta\hat{T}}{\hat{\rho}_0\hat{R}_g\hat{T}_g\hat{H}_0}$ (measure of Kelvin effect), $\psi=\frac{\hat{M}\hat{L}\Delta\hat{T}}{\hat{R}_g\hat{T}_g^2}$ (effect of local temperature difference on the mass flux), and Jakob number, $\mathrm{Ja}=\frac{\hat{k}_0\Delta\hat{T}}{\hat{H}_0\hat{L}\hat{p}_{v,sat}}\sqrt{\frac{2\pi\hat{R}_g\hat{T}_g}{\hat{M}}}$, measuring the importance of kinetic effects at the interface. 

For sessile droplets with slow interfacial phase change, the inertia effect can be neglected with very small Reynolds number (usually Re $<10^{-1}$ for droplet wetting and evaporation), therefore we set the Reynolds number to zero in the formulation. By further ignoring the parameter variation in the azimuthal direction, we arrive at the dimensionless governing equations, including the conservation equations of mass, Eq. (\ref{eq:mass_scaled}), momentum, Eq. (\ref{eq:momentum_scaled}) and energy, Eq. (\ref{eq:energy_scaled}), as well as the dimensionless boundary equations, including the normal and tangential stress boundary balance, Eq. (\ref{eq:stress_boundary_scaled}), the kinematic boundary condition, Eq. (\ref{eq:kinematic_boundary_scaled}), the jump energy balance, Eq. (\ref{eq:jump_energy_scaled}), and the expression of interfacial mass flux, Eq. (\ref{eq:mass_flux_scaled}). 
\begin{equation}
    \frac{1}{r}\frac{\partial(ru)}{\partial r}+\frac{\partial w}{\partial z}=0,
    \label{eq:mass_scaled}
\end{equation}
\begin{equation}
    \frac{\partial p}{\partial r}=\frac{\partial}{\partial z}(\mu\frac{\partial u}{\partial z}),
    \label{eq:momentum_scaled}
\end{equation}
\begin{equation}
    \frac{\partial}{\partial z}(k \frac{\partial T}{\partial z})=0,
    \label{eq:energy_scaled}
\end{equation}
\begin{equation}
    p=-\frac{\varepsilon^2\sigma}{Ma}(\frac{1}{r}\frac{\partial}{\partial r}(r\frac{\partial h}{\partial r}))-\frac{\mathcal{A}}{h^3},\ \tau_{zr}=\frac{1}{Ma}\frac{\partial\sigma}{\partial r},
    \label{eq:stress_boundary_scaled}
\end{equation}
\begin{equation}
    \frac{\partial h}{\partial t}+u\frac{\partial h}{\partial r}-w+EJ=0,
    \label{eq:kinematic_boundary_scaled}
\end{equation}
\begin{equation}
    J+k\frac{\partial T}{\partial z}=0,
    \label{eq:jump_energy_scaled}
\end{equation}
\begin{equation}
    \mathrm{Ja}J=\delta p+\psi (T_i-T_g)+\mathrm{ln}(\frac{1}{\chi_\mathrm{vapor}}),
    \label{eq:mass_flux_scaled}
\end{equation}

The above equations are integrated over the vertical direction, with the integration of velocity and temperature expressed as $f=\int_0^h udz,\ \Theta=\int_0^h Tdz$, and the governing equations are subsequently discretized using a finite element/Galerkin method; the variables are approximated using quadratic Lagrangian basis functions $\Phi_i$. After applying the divergence theorem, we arrive at the weak form of five governing equations, Eqs. (\ref{eq:mass_scaled_int}-\ref{eq:mass_flux_scaled_int}), with five independent variables, $h$, $p$, $f$, $\Theta$, and $J$.
\begin{equation}
    R=\int(\frac{\partial h}{\partial t}\Phi_i+EJ\Phi_i-f\frac{\partial\Phi_i}{\partial r})rdr+(rf\Phi_i)_{r=0}^{r=r_\infty},
    \label{eq:mass_scaled_int}
\end{equation}
\begin{equation}
    R=\int(h\frac{\partial p}{\partial r}-(\mu\frac{\partial u}{\partial z})_0^h)\Phi_i rdr,
    \label{eq:momentum_scaled_int}
\end{equation}
\begin{equation}
    R=\int(k \frac{\partial T}{\partial z})_0^h\Phi_i rdr,
    \label{eq:energy_scaled_int}
\end{equation}
\begin{equation}
    R=\int((p+\frac{\mathcal{A}}{h^3})\frac{Ma}{\varepsilon^2\sigma}\Phi_i-\frac{\partial h}{\partial r}\frac{\partial \Phi_i}{\partial r})rdr+(r\frac{\partial h}{\partial r}\Phi_i)_{r=0}^{r=r_\infty},
    \label{eq:stress_boundary_scaled_int}
\end{equation}
\begin{equation}
    R=\int(\mathrm{Ja}J-\delta p-\psi(T|_{z=h}-T_g)+\mathrm{ln}\chi_\mathrm{vapor})\Phi_i rdr.
    \label{eq:mass_flux_scaled_int}
\end{equation}

Finally, we consider an Ansatz distribution function of liquid temperature \emph{T} and velocity \emph{u} in \emph{z} axis to complete the mathematical formulation. We fit the flow velocity and the temperature distribution with a parabolic expression, \emph{i.e.} $y=c_3+c_2z+c_1 z^2$, then derive their expressions by applying the boundary conditions at $z = 0$ and $z = h$.

At the liquid-gas interface, $z = h$, the flow velocity meets the tangential stress balance Eq. (\ref{eq:stress_boundary_scaled}), which relates the surface tension gradient to the shear stress (multiplication of dynamic viscosity and velocity gradient). At the liquid-solid interface, $z = 0$, a no slip boundary condition applies. The flow velocity is thereafter derived as,
\begin{equation}
    u=(\frac{3f}{h^2}-\frac{1}{2\mu \mathrm{Ma}}\frac{\partial\sigma}{\partial r})z+(\frac{3}{4\mu h \mathrm{Ma}}\frac{\partial\sigma}{\partial r}-\frac{3f}{2h^3})z^2.
\end{equation}

For the temperature distribution, the temperature meets the jump energy balance at the droplet surface, $z = h$, Eq. \ref{eq:jump_energy_scaled}, which relates the temperature gradient in the liquid side to the interfacial heat flux due to phase change. At the liquid-solid interface, we consider substrates both with constant temperature and with downward heat dissipation. For cases where the substrate temperature is kept constant with temperature controller, the temperature distribution in the liquid side is derived as,
\begin{equation}
    T=T_W+(\frac{3\Theta}{h^2}-\frac{3T_W}{h}+\frac{J}{2k})z+(\frac{3T_W}{2h^2}-\frac{3\Theta}{2h^3}-\frac{3J}{4kh})z^2.
\end{equation}

For a thermal conductive substrate with finite thickness $H_{\rm solid}$ and thermal conductivity $k_{\rm solid}$, the process can be simplified into a quasi-steady one-dimensional problem. Similar approximations were also adopted in the studies of \cite{ristenpart2007influence}, \cite{xu2010criterion} and \cite{dunn2009strong}. At small Reynold numbers, the heat transfer in the vertical direction is dominated by heat conduction rather than convection. In such cases, the boundary condition at $z = 0$ in this situation can be described as,
\begin{equation}
    k\frac{\partial T}{\partial z}=\frac{T-T_W}{\rm Bi},
\end{equation}
where Biot number, $\mathrm{Bi}= H_{\mathrm{solid}}⁄k_\mathrm{solid}=\hat{H}_{\mathrm{solid}}\hat{k}_0⁄\hat{H}_0\hat{k}_\mathrm{solid}$, denotes the ratio of thermal resistances inside the solid phase and the liquid phase. The distribution of liquid temperature in such cases is thereafter derived as,
\begin{equation}
\begin{split}
    T=c_3+c_2z+c_1z^2, \mathrm{where}\ \\
    c_1=\frac{1}{3\mathrm{Bi}k+h}(-\frac{3\mathrm{Bi}J}{2h}-\frac{3J}{4k}-\frac{3\Theta}{2h^2}+\frac{3T_W}{2h}), \\
    c_2=\frac{1}{3\mathrm{Bi}k+h}(\frac{hJ}{2k}+\frac{3\Theta}{h}-3T_W),\\
    c_3=\frac{1}{3\mathrm{Bi}k+h}(\frac{\mathrm{Bi}hJ}{2}+\frac{3\mathrm{Bi}k\Theta}{h}+hT_W).
\end{split}
\end{equation}

\vspace{20pt}
\textbf{Initial and Boundary Conditions} In the region of $0\leq r\leq 1$, the initial conditions, $t=0$, are set as,
\begin{equation}
    h(r,0)=h_\infty+1-r^2,\ f(r,0)=0,\ \Theta(r,0)=h(r,0)T_W,
\end{equation}
where $h_\infty$ is the thickness of the precursor film, $T_W$ is the initial temperature of the solid substrate. For zero mass flux at the precursor region, the initial value of $h_\infty$ is set as $\sqrt[3]{\frac{\mathcal{A}\delta}{\psi(T_W-T_g)-\mathrm{ln}(\chi_\mathrm{vapor})}}$. The parameters at the droplet center, $r = 0$, satisfy the symmetric boundary conditions, expressed as,
\begin{equation}
    \frac{\partial h}{\partial r}(0,t)=0,\ f(0,t)=0,\ \frac{\partial\Theta}{\partial r}(0,t)=0.
\end{equation}

In the region of precursor film, $r > 1$, the parameters are set as,
\begin{equation}
    h(r,0)=h_\infty,\ f(r,0)=0,\ \Theta(r,0)=h_\infty T_W.
\end{equation}

We apply finite element Galerkin method to solve the system of governing equations. One dimensional mesh is constructed along the \emph{r}-direction, and the length of the computational domain is set properly according to the actual situation of each condition to ensure the free development of the droplet morphology. The domain is discretized along \emph{r} from 0 to $r_\infty$ into equally spaced nodes, the total number being $N_\mathrm{tot}$, with grid independence test of the solution. At each time step, we apply the Newton-Raphson method to obtain the solution across the computational domain progressively. The solution evolves forward with an adaptive time interval which adjusts according to the maximum residual errors of the governing equations from the previous time step.

\section{Experimental Method}\label{sec:experimental_method}

We use high speed camera (Photron FASTCAM Mini AX200 type with frame rate of 3,000 fps) to trace the fast spreading of droplet upon its contact with a solid surface. To avoid the influence of initial impinging force of the droplet, we set the distance between the syringe and the substrate to a similar value with the initial diameter of the spherical droplet ($V_0 =3\pm0.2\mu L$ for 1-Butanol and 2-Propanol, and $0.6\pm 0.05\mu L$ for FC-72), so that the droplet can touch the solid surface right after it is generated from the syringe tip. We selected 1-Butanol, 2-Propanol (IPA) and FC-72 as the working fluids, each with one order of magnitude larger saturation vapor pressure (Table \ref{tab:liquid_properties}). Three types of solid materials are utilized, \emph{i.e.}, copper, SUS430 and glass, each with one order of magnitude smaller thermal conductivity. We also chose copper substrates of thickness 0.3mm, 1mm and 3mm to demonstrate the effect of substrate thickness. The three types of substrates exhibit a highly wetting state for 1-Butanol, 2-Propanol and FC-72 which matches the requirement of low contact angle (less than 40°) for lubrication approximation. For each condition, we repeated the experiments for at least 5 times. The average values of the spreading rates along with standard deviations are derived for comparison with the numerical predictions (Table \ref{tab:num_vs_exp}). The key thermophysical properties of different liquids and solids are given in Table \ref{tab:liquid_properties} and \ref{tab:solid_conductivity}.

\begin{table}
  \begin{center}
\def~{\hphantom{0}}
  \begin{tabular}{ccccccccc}
      Liquid&\makecell{$\hat{\rho}$ \\($kg/m^3$)}&\makecell{$\hat{\mu}$ \\($mPa\cdot s$)}&\makecell{$\hat{\sigma}$ \\($mN/m$)}&\makecell{$\hat{\eta_\sigma}$ \\($mN/m/K$)}&\makecell{$\hat{k}$\\ ($W/m/K$)}&\makecell{ $\hat{c}_p$\\($J/kg/K$)}&\makecell{$\hat{p}_\mathrm{sat}$\\(Pa)}&\makecell{$\hat{L}$\\ ($kJ/kg$)} \\
Water & 998 & 1.005 & 72.75 & -0.15 & 0.60 & 4180 & 2341 & 2257\\
1-Butanol& 810& 2.573& 24.57&	-0.07&	0.1553&	2386&	580&	584\\
2-Propanol&	786&	2.012&	23.0&	-0.09&	0.135&	1527&	4420&	663\\
FC-72&	1680&	0.64&	10&	-& 0.057&	1100&	30900&	88\\
Methanol&	792&	0.59&	22.7&	-0.0773&	0.204&	2460&	10300&	1191\\
Ethanol&	789&	1.144&	22&	-0.0832&	0.166&	2570&	5786&	1030 \\
Chloroform&	1490&	0.563&	27.5&	-0.1295&	0.13&	957&	25900&	248 \\
Heptane&	684&	0.389&	20.2&	-0.098&	0.128&	2220&	4621&	3553 \\
Hexane&	659&	0.301&	18.4&	-0.1022&	0.125&	2243&	19140&	382

  \end{tabular}
  \caption{Thermophysical properties of representative liquids at 293 K except FC-72 (298K). 1-Butanol, 2-Propanol and FC-72 are utilized in present experiments. Water, 2-Propanol, Methanol, Ethanol and Chloroform are utilized in previous studies on circulation reversal (\cite{ristenpart2007influence}, \cite{xu2010criterion}, \cite{li2015coffee}) as marked in Fig. \ref{fig:flow_reversal}. Alkanes (e.g., Heptane and Hexane) are utilized in the study of \cite{shiri2021thermal} on the influence of thermal Marangoni flow on droplet geometry as marked in Fig. \ref{fig:spreading_law}.}
  \label{tab:liquid_properties}
  \end{center}
\end{table}
\begin{table}
  \begin{center}
\def~{\hphantom{0}}
  \begin{tabular}{ccccccccc}
    Material Type&	Copper&	SUS430&	Glass&	Polydimethylsiloxane (PDMS)\\
$k_\mathrm{solid}$ ($W/m/K$)&	385&	25&	1.2&	0.23

  \end{tabular}
  \caption{Thermal conductivity of representative substrate materials. Copper, SUS430 and glass are utilized in present study. Glass is utilized in the work of \cite{xu2010criterion} and \cite{hu2006marangoni}. PDMS is utilized in the work of \cite{ristenpart2007influence}.}
  \label{tab:solid_conductivity}
  \end{center}
\end{table}

To remove any oxide residuals and ensure the uniformity and smoothness of the solid surface, we fabricate substrate surfaces with mirror finishing. The surface roughness of glass, SUS430 and copper is characterized to be $\mathrm{Sa}_\mathrm{glass} = 0.258 \mu m$, $\mathrm{Sa}_\mathrm{SUS430} = 0.134 \mu m$, $\mathrm{Sa}_\mathrm{copper} = 0.291 \mu m$ with 3D laser scanning microscope (OLYMPUS/LEXT OLS4000). Before each set of experiments, we prepare the solid substrate by 15 min ultrasonic bath with 99.8$\%$ ethanol, then flush with large quantity of deionized water, and dry the surface with high speed clean air flow. We start the high speed camera right before the deposition of the droplet and stop recording as the droplet fully spreads. The videos are further processed with imageJ (\cite{schneider2012nih}) and self-developed Matlab codes, where the start of the measurement is set as the moment when a droplet fully detaches from the syringe tip and forms a spherical cap at the solid surface.

Microparticle image velocimetry ($\mu$PIV) is implemented to measure the flow field near the bottom of an evaporating droplet. The measurements are based on an inverted fluorescent microscope (Nikon Ts2-FL). Specifically, a green laser (468nm) is utilized to excite the fluorescent micro tracers (Fluoro-Max; green fluorescent polymer microspheres: excitation/emission 468/508 nm, diameter 1.0 $\mu$m). A CCD camera (STC-MCS510U3V) connected to the microlens captures the fluorescent signal emitted by the tracer particles. We use butanol and IPA as the test fluids, which have similar surface tension but different volatility (Table \ref{tab:liquid_properties}). Experiments are conducted on slide glass for microscope with thickness of $1.1 \pm0.2$ mm. The environmental condition is controlled as 25°C and 50$\%$ RH during the experiments. The droplet size for microPIV experiments is set as $0.5 \pm 0.002\mu L$, and is focused with 20x lens from the bottom. Each set of experiments is repeated for more than five times. The representative videos are presented as in Supplementary Movie 3, where the video of Butanol is speeded up by 20 times and the video of IPA is slowed down by 3 times with ImageJ.
 
\section{Results and Discussion}\label{sec:results_discussion}
\subsection{Transition of flow states – physical mechanisms} 
\begin{figure}
    \centering
    \includegraphics[width=10cm]{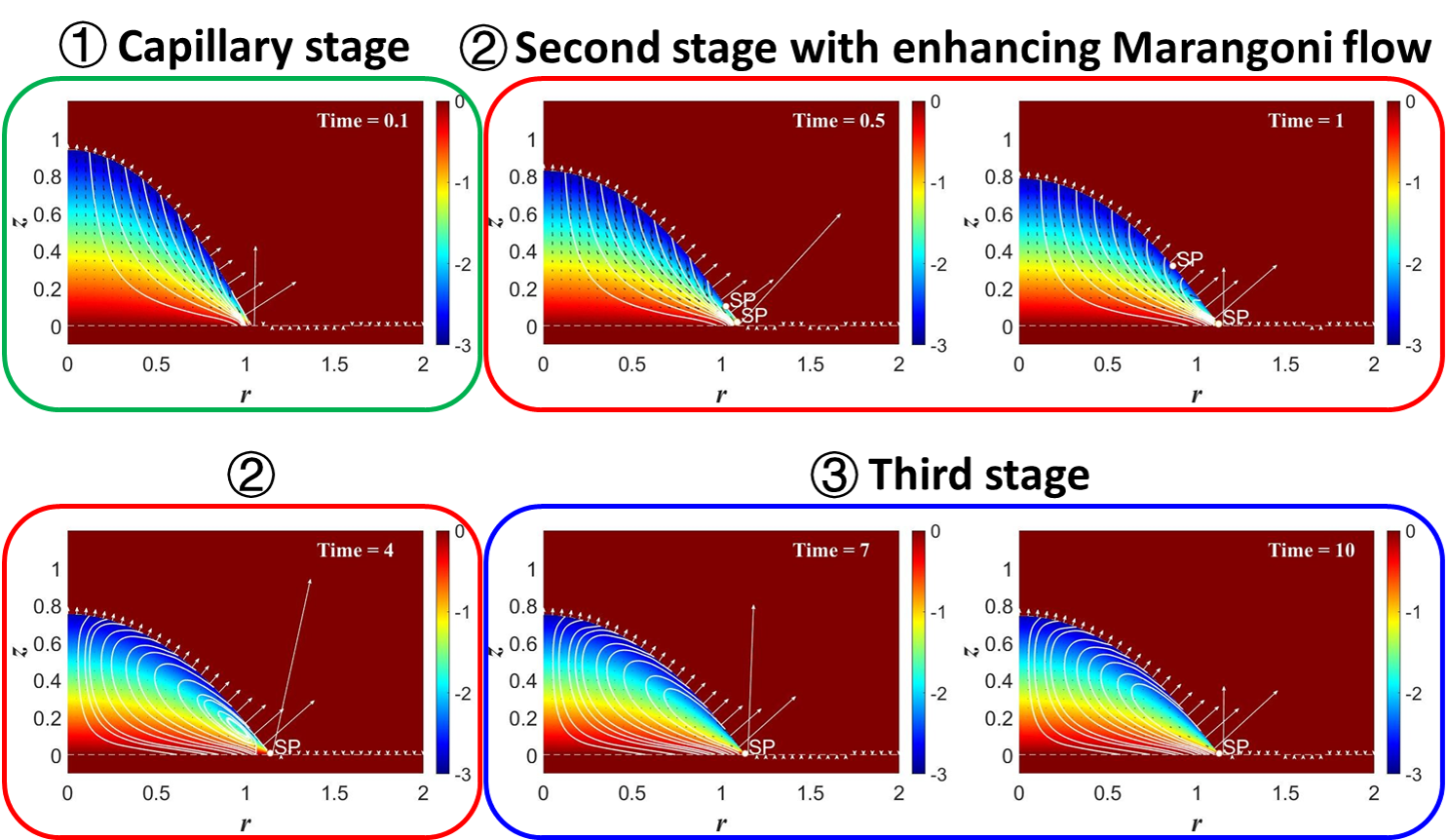}
    \caption{Evolution of flow state for an evaporative droplet on a thermal conductive substrate. The three stages are classified according to the state of internal flow, where continuous outward capillary flow dominates droplet spreading at the initial stage, the Marangoni flow develops at the second stage, and balances with the capillary flow at the third stage. The key dimensionless numbers in this case are set as $\rm Ja = 5\times 10^{-3}$, $\rm Bi = 10^{-4}$, $\rm Ma = 10^{-2}$.}
    \label{fig:flow_field}
\end{figure}

Fig. \ref{fig:flow_field} shows the evolution of flow field in a representative case, which can be classified into three stages. At the initial stage ($\textcircled{\scriptsize{1}}$) when a liquid droplet contacts a solid surface, the capillary effect drives a continuous outward flow, leading to rapid droplet spreading. In the second stage ($\textcircled{\scriptsize{2}}$), the evaporation cooling effect and the nonuniform heat dissipation results in temperature gradient across the droplet surface, and induces thermal Marangoni flow from the droplet edge toward the apex. The opposing thermal Marangoni and capillary forces lead to the formation of two stagnation points (SP) at the droplet surface. The upper stagnation point moves towards the droplet apex as the Marangoni flow gets stronger and disappears with a recirculation vortex that engulfs the whole droplet. The lower stagnation point remains near the contact line ($\textcircled{\scriptsize{3}}$) as a result of capillary-Marangoni interaction (\cite{xu2007marangoni}, \cite{ristenpart2007influence}, \cite{wang2018stagnation}). 

Yet two questions arise as our analysis goes on. One is ‘does the stagnation point universally exist?’ Another is ‘does thermal Marangoni flow always play a role in wetting dynamics?’ These two questions are intricately related as the relative strength of thermal Marangoni flow and capillary flow decides the flow state and the formation of the stagnation point, and therefore the wetting dynamics.

To elucidate this, we conducted detailed parametric analyses on all possible combinations of liquids and solids. Two dominant dimensionless numbers are extracted which govern the interacting physics, namely, the Jakob number, Ja, which evaluates the balance between interfacial phase change and heat transfer into the liquid phase, and the liquid-solid Biot number, Bi, which indicates the ratio of liquid phase to solid phase thermal resistance.
\begin{equation}
\mathrm{Ja}=\frac{\hat{k}_0\Delta \hat{T}}{\hat{H}_0 \hat{L}\hat{p}_{\text{v,sat}}} \sqrt{\frac{2\pi\hat{R}_\text{g} \hat{T}_\text{g}}{\hat{M}}},\  \mathrm{Bi}=\frac{\hat{H}_\text{solid}\hat{k}_0}{\hat{H}_0\hat{k}_\text{solid}},
\end{equation}
where $\hat{k}_0$ is the thermal conductivity of the liquid, $\Delta T$ is the maximum temperature difference across the droplet, $\hat{H}_0$ is the initial droplet height, $\hat{L}$ is the latent heat of the liquid, $\hat{M}$ is the molar mass of the liquid, $\hat{p}_\mathrm{v,sat}$ is the saturation vapor pressure of the liquid, $\hat{R}_\mathrm{g}$ is the gas constant, $\hat{T}_\mathrm{g}$  is the gas phase temperature, $\hat{H}_\mathrm{solid}$ is the substrate thickness, and $\hat{k}_\mathrm{solid}$ is the substrate thermal conductivity. 

With decreasing Bi, heat dissipation into the substrate becomes efficient. In such cases, thermal resistance in the liquid phase dominates, thus the temperature gradient across the droplet surface is large from the droplet apex to the edge (large difference in thermal resistance across the liquid). The large temperature gradient strengthens the Marangoni flow and retards droplet spreading.

The Jakob number, mostly related to liquid volatility, decides the strength of evaporation cooling. For moderately volatile liquids, the evaporation mass flux takes heat away from the droplet surface, while the heat supply from the substrate is also considerable. Near the contact line, the liquid film is thin, enabling good heat supply and therefore a higher temperature in comparison to the droplet apex; this is the case for many volatile liquids such as water and alcohol. At very high liquid volatility (small Ja, \emph{e.g.} flash evaporation), heat supply from the substrate can no longer meet the heat loss from the droplet surface, as a result, the droplet surface gets ‘uniformly cold’, leading to weak thermal Marangoni flow. For liquids that hardly evaporate (low volatility and large Ja), the effect of evaporation cooling is negligible, and the process is near isothermal, which also leads to weak Marangoni effect. 

\begin{figure}
    \centering
\includegraphics[width=10cm]{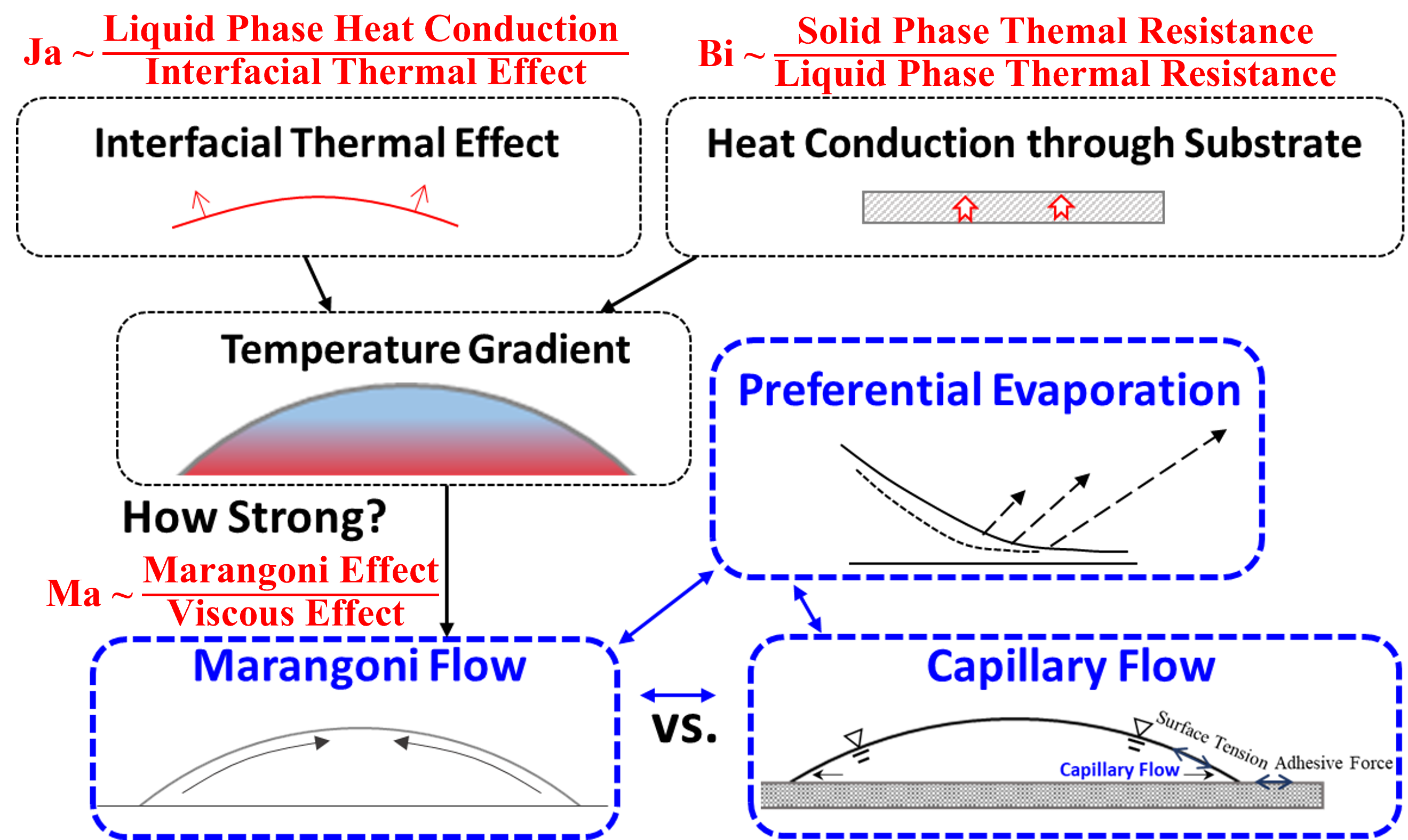}
\caption{Decomposition of the dominating mechanisms in wetting dynamics of evaporative droplets. }
 \label{fig:dominating_mechanisms}
\end{figure}

Taking an overview of the dominating mechanisms, we can see that Ja (interfacial thermal effect) and Bi (heat conduction into the substrate) determine the temperature field in the droplet-solid system. Once the temperature gradient establishes, the Marangoni number determines the strength of the Marangoni flow, defined as $\rm Ma=\frac{\hat{\mu}_0 \hat{\emph{u}}^*}{\varepsilon \hat{\sigma}_0}=\frac{\hat{\eta}_\sigma \Delta\hat{\emph{T}}}{\hat{\sigma}_0}$, where the characteristic velocity $\hat{u}^*$ is defined as $\hat{u}^*=\frac{\varepsilon \hat{\eta}_\sigma \Delta\hat{T}}{\hat{\mu}_0}$, $\hat{\mu}_0$ denotes the liquid viscosity, $\hat{\sigma}_0$ is the liquid-gas surface tension at standard conditions, $\varepsilon$ is the aspect ratio of the droplet, $\hat{\eta}_\sigma$ is the coefficient of surface tension to temperature, and $\Delta\hat{T}$ is the maximum temperature difference across the droplet. The Marangoni flow interacts with the capillary flow, which together affect the flow field inside the droplet, and further determine the droplet dynamics along with the nonuniform distribution of evaporation mass flux, as indicated by Fig. \ref{fig:dominating_mechanisms}.

\subsection{Conditions for flow reversal} 
To quantify the role of different effects on droplet dynamics, we decompose the dimensionless interfacial flow velocity $u_\mathrm{s}$ into thermal Marangoni velocity, $u_\mathrm{tg}$, resulting from interfacial temperature gradient, and the capillary velocity, $u_\mathrm{ca}$, related to the local surface curvature. 
\begin{equation} 
\begin{split}
u_\text{s} = -\frac{h^2}{2\mu} \frac{\partial p}{\partial r} - \frac{h}{\mu} \frac{\partial T_\text{s}}{\partial r},\ u_{\text{tg}}=\frac{h}{\mu} \frac{\partial T_s}{\partial r},\ u_{\text{ca}}=-\frac{h^2}{2\mu} \frac{\partial p}{\partial r},
\end{split}
\end{equation}
where \emph{h} is the local height of the liquid film, $\mu$ is the liquid viscosity, \emph{p} is the pressure at the liquid phase, \emph{r} is the coordinate in the radial direction, $T_\mathrm{s}$ is the interfacial temperature. 

\begin{figure}
    \centering
    \includegraphics[width=13cm]{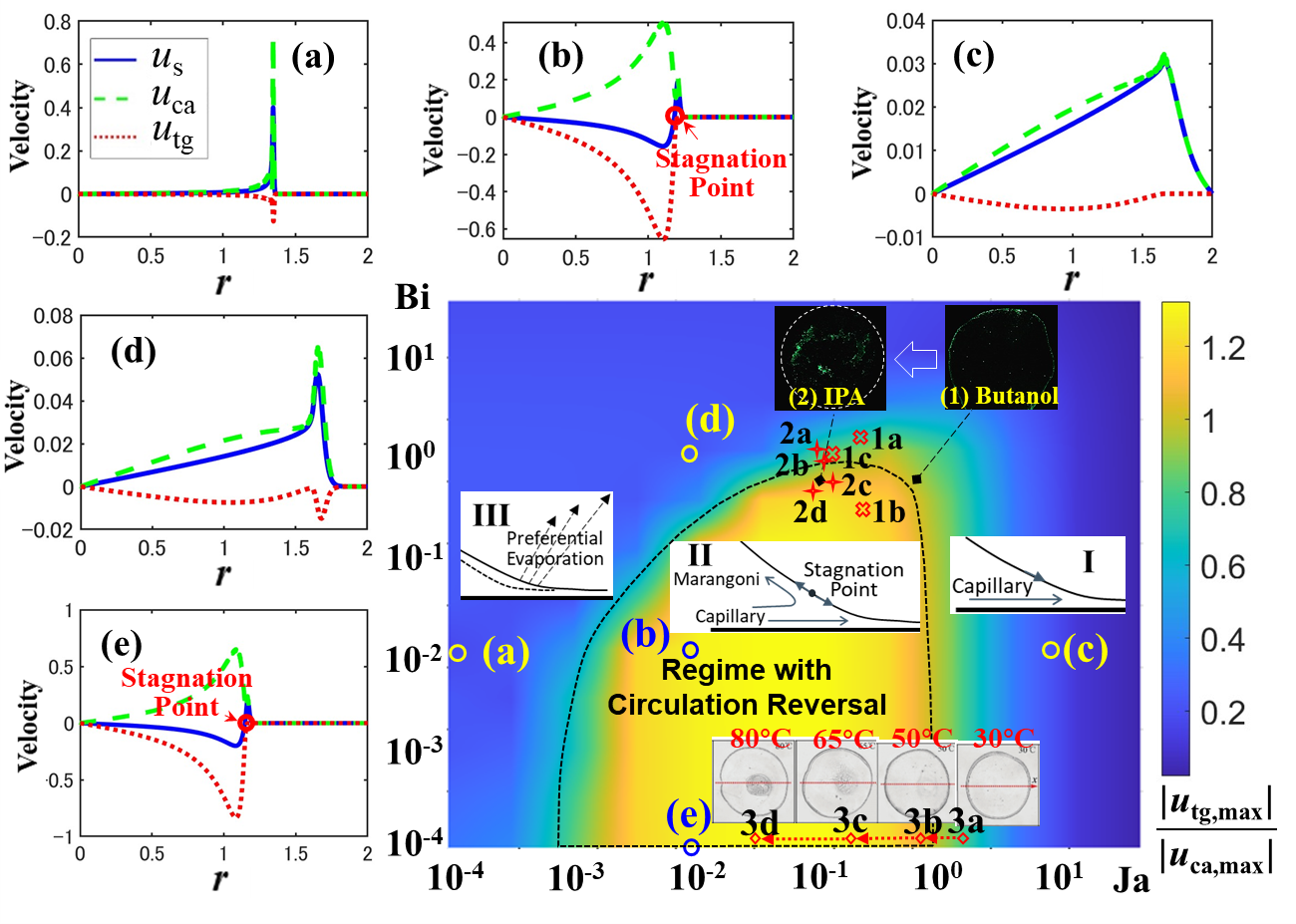}
    \caption{Phase diagram for the relative strength of thermal Marangoni flow to capillary flow. Figures (a) to (e) show the decomposition of surface velocity $u_\mathrm{s}$ to thermal Marangoni velocity $u_\mathrm{tg}$ and capillary velocity $u_\mathrm{ca}$, corresponding to letters (a) to (e) in the colormap. Sketches I, II, III indicate the dominating mechanism at different regimes of the phase diagram. The experimental data by \cite{xu2010criterion} (1a: water on glass of thickness 1 mm, 1b: water on glass of thickness 0.15 mm, 1c: IPA on glass of thickness 1 mm), \cite{ristenpart2007influence} (2 $\mu$L droplets of 2a: Methanol, 2b: Ethanol, 2c: IPA, 2d: Chloroform, on PDMS of thickness 4mm) and \cite{li2015coffee} (3a to 3b with leftward dotted arrows and representative deposition patterns: 2.5 $\mu$L droplets on glass substrate of 30, 50, 65, and 80 $^\circ \rm C$) are marked out, which indicate the flow reversal inside the droplet with changing environmental settings. Note that the dotted line presents the criterion for the formation of stagnation point, and does not necessarily correspond to the contour line of $|u_\mathrm{tg,max}|/|u_\mathrm{ca,max}|=1$.}
    \label{fig:flow_reversal}
\end{figure}

Fig. \ref{fig:flow_reversal} maps the relative strength of thermal Marangoni flow and capillary flow, $| u_\mathrm{tg,max}|/|u_\mathrm{ca,max}|$, for a wide range of combinations of Ja and Bi, along with decomposed interfacial velocity (a)$\sim$(e). With decreasing Bi from (d) to (b) to (e), the thermal Marangoni flow enhances and outweighs the capillary effect. As a result, the direction of interfacial flow reverses at a position close to the contact line, forming a stagnation point as marked in Fig. \ref{fig:flow_reversal}(b) and (e). 

From (a) to (b) to (c), the value of Ja increases, corresponding to decreasing liquid volatility. At a small value of Ja, severe liquid-vapor phase change takes place, preferentially at the contact line region, leading to the concentrated interplay of Marangoni and capillary effects near the contact line (Fig. \ref{fig:flow_reversal}(a)). The severe interfacial phase change also leads to uniformly cold droplet surface, corresponding to weak Marangoni effect. At large value of Ja (Fig. \ref{fig:flow_reversal}(c)), the whole process is close to isothermal due to weak evaporation. In such cases, the thermal Marangoni flow is also weak, and the droplet spreading is dominated by capillary flow, close to the case described by Tanner’s law.

Overall, the thermal Marangoni effect becomes apparent and even outweighs the capillary effect at limited combinations of Ja and Bi. In regime II circled by dot lines in Fig. \ref{fig:flow_reversal}, the thermal Marangoni effect is strong enough to reverse the flow direction and form a stagnation point at the droplet surface. The droplet dynamics within this regime is therefore a joint result of the thermal Marangoni effect, evaporation effect, and capillary effect, in contrast to the high-Ja regime where capillary effect dominates (Regime I) and the low-Ja regime where liquid depletion due to evaporation near the contact line plays a significant role (Regime III).

Previous studies revealed the circulation reversal in evaporating droplets with varying substrate conductivity. We recalculated the experimental data in these pioneer work (\cite{ristenpart2007influence}, \cite{xu2010criterion}), which are demonstrated in the phase diagram and show good correspondence with our theoretical predictions. Due to technical limitations, liquids in those studies are limited to water and alcohols where the evaporation is moderate with sufficient time for observation, and the substrates are mostly glass for microPIV visualization. With the criterion proposed in Fig. \ref{fig:flow_reversal}, we are able to quickly find out the flow state for liquids with varying thermal conductivity and volatility, and on substrates with varying thickness and thermal properties, which could not have been possible by experimental observations. 

\begin{figure}
    \centering
    \includegraphics[width=12cm]{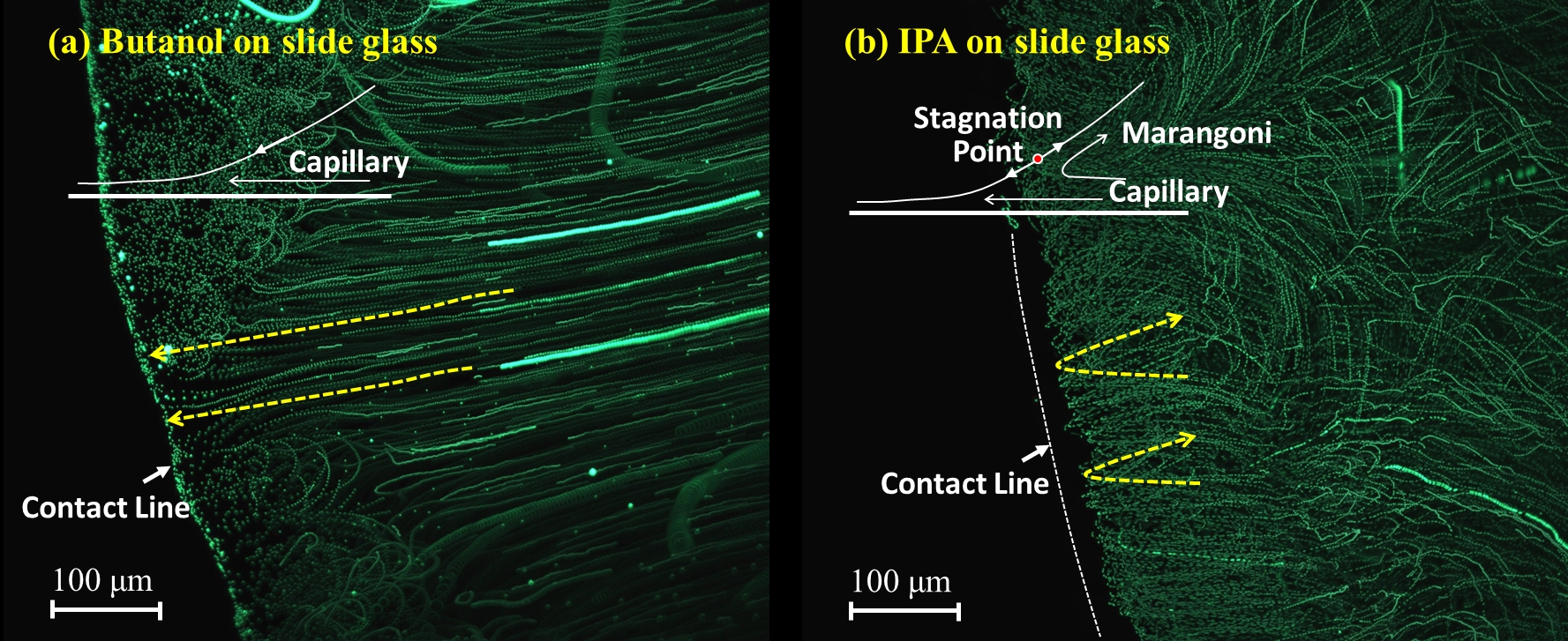}
    \caption{Flow pattern near three-phase contact line visualized by MicroPIV where the tracing tracks of fluorescent particles are overlap of 100 continuous frames: (a) a 0.5$\mu L$ Butanol droplet on slide glass, and (b) a 0.5$\mu L$ IPA droplet on slide glass.}
    \label{fig:microPIV}
\end{figure}

Corresponding to the flow state inside the droplet, the deposition patterns from drying droplets differ, which provides a possible way for controllable pattern formation in applications including printing, coating and thin film fabrication. As indicated in Fig. \ref{fig:flow_reversal}, the deposition patterns transit from coffee ring (\cite{deegan1997capillary}) to coffee eye with increasing liquid volatility (\cite{hu2006marangoni}) and increasing substrate temperature (\cite{li2015coffee}) where the thermal Marangoni flow strengthens, reverses the flow direction, and leads to the formation of coffee eye at the center as indicated by diamond marks 3a to 3d in Fig. \ref{fig:flow_reversal}. Further visualizations with microPIV also demonstrate the flow reversal as the liquid changes from low volatility Butanol to high volatility IPA on slide glass (Fig. \ref{fig:microPIV}, Supplementary Movie 3). In correspondence to the flow direction, the deposition transits from coffee-ring to more uniform patterns as indicated by the deposited fluorescent tracing particles suspended in (1) Butanol and (2) IPA droplets in Fig. \ref{fig:flow_reversal}.

\subsection{Spreading law of volatile droplets} 
\begin{figure}
    \centering
    \includegraphics[width=13cm]{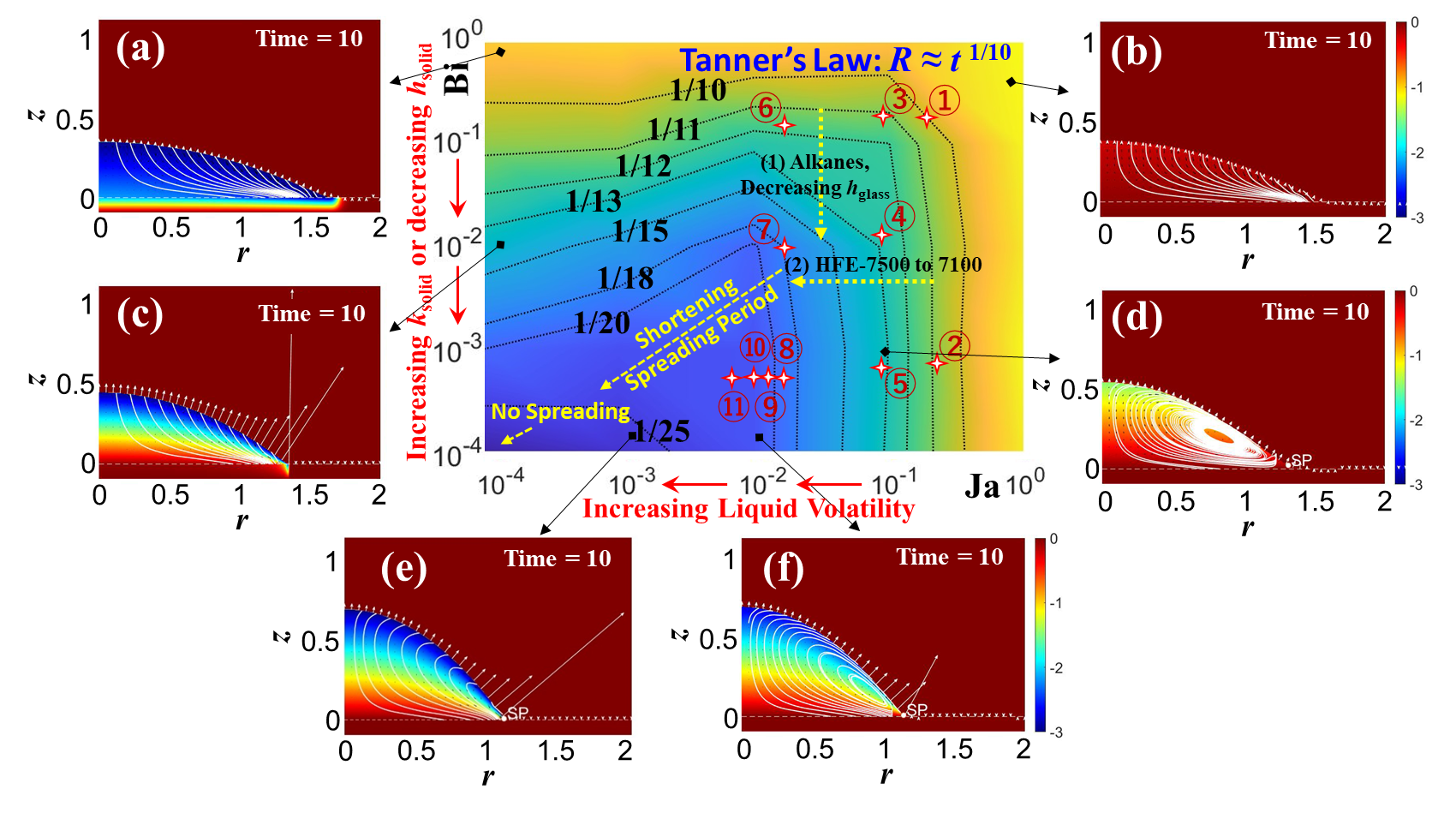}
    \caption{Phase diagram for the spreading rate of droplets along with demonstrations of flow and temperature fields in representative conditions. Contour lines of the spreading exponent are marked out with corresponding values. Experimental results by \cite{shiri2021thermal} ((1) Alkanes on glass substrates with decreasing thickness) and \cite{kumar2022wetting} ((2) HFE 7500 to 7100 with increasing volatility) are presented with yellow dotted arrows. Experimental results in our research are marked with stars numbered as $\textcircled{\scriptsize{1}}$ Butanol on 1.1 mm glass, $\textcircled{\scriptsize{2}}$ Butanol on 1.0 mm copper, $\textcircled{\scriptsize{3}}$ IPA on 1.1mm glass, $\textcircled{\scriptsize{4}}$ IPA on 1.0mm SUS430, $\textcircled{\scriptsize{5}}$ IPA on 1.0mm copper, $\textcircled{\scriptsize{6}}$ FC-72 on 1.1mm glass, $\textcircled{\scriptsize{7}}$ FC-72 on 1.0mm SUS430, $\textcircled{\scriptsize{8}}$ FC-72 on 1.0mm copper ($\textcircled{\scriptsize{1}}\sim\textcircled{\scriptsize{8}}$ are cases with substrate temperature $T_{\rm sub}$= 20$^\circ \rm C$), $\textcircled{\scriptsize{9}}$ FC-72 on 1.0mm copper ($T_{\rm sub}$ = 25$^\circ \rm C$), $\textcircled{\scriptsize{10}}$ FC-72 on 1.0mm copper ($T_{\rm sub}$= 35$^\circ \rm C$) and $\textcircled{\scriptsize{11}}$ FC-72 on 1.0mm copper ($T_{\rm sub}$ = 45$^\circ \rm C$). }
    \label{fig:spreading_law}
\end{figure}

The interfacial phase change and interacting flows further decide the spreading dynamics of the droplet. Specifically, the speed of contact line can be quantified with a power law in the form of $R(t) = t^n$, where \textit{n} is the spreading exponent. We then check the flow state and evaluate the corresponding spreading rate for a wide range of Ja and Bi. A phase diagram is thereafter derived with representative cases shown in Fig. \ref{fig:spreading_law}. Overall, the spreading exponent decreases from 1/10 (Tanner’s law) to smaller values (1/11, 1/12, … 1/25) with decreasing Ja and decreasing Bi. As Bi decreases, heat transport across the substrate gets efficient and the thermal resistance in the liquid phase becomes dominant. This strengthens both the evaporation mass flux and the thermal Marangoni flow (enlarged temperature gradient).

With decreasing Ja, the evaporation effect enhances, while the thermal Marangoni flow strengthens first (from isothermal state to large temperature gradient) and then weakens (as the droplet surface gets uniformly cold due to strong evaporation cooling effect). Overall, the influence of substrate thermal properties on the spreading rate is weak at small liquid volatility and gets significant at high liquid volatility as the evaporation cooling effect and the thermal transport process becomes eminent. These conclusions allow for optimal design of the thermal properties of working fluids and substrates, where a balance between liquid wetting and thermal efficiency is needed for specific application scenarios, \emph{e.g.}, in avoiding transition from dropwise to filmwise condensation, or in avoiding dry-out in cooling of high-power-density electronics.

To further validate our theoretical findings, a series of experiments are conducted with selected liquids and solid substrates. To ensure uniform spreading and minimize hysteresis, we prepare substrates with mirror finishing (roughness characterized as $\rm Sa_{glass} = 0.258\ \mu m$, $\rm Sa_{SUS430} = 0.134\ \mu m$, $\rm Sa_{copper} = 0.291\ \mu m$), and the substrate temperature is controlled as $20\pm0.2\ ^\circ \rm C$ with a Peltier module. Experiments are conducted with liquids of different magnitudes of volatility, \emph{i.e.} 1-Butanol ($p_{\rm sat,20^\circ C} = 580\ \rm Pa$), 2-Propanol (IPA, $p_{\rm sat,20^\circ C} = 4420\ \rm Pa$) and FluorinertR (FC-72, $p_{\rm sat,25^\circ C} = 30900\ \rm Pa$) and on substrates with stepwise varying thermal conductivity and thickness, \emph{i.e.} glass ($k_{\rm glass} = 0.8\ \rm W/m/K$, $h_{\rm glass} = 1.1\ \rm mm$), SUS430 ($k_{\rm SUS430} = 25\ \rm W/m/K$, $h_{\rm SUS430} =1.0\ \rm mm$), copper ($k_{\rm copper} = 398\ \rm W/m/K$, $h_{\rm copper} = 0.1,\ 1.0,\ 3.0\ mm$). The results are marked with numbered stars in Fig. \ref{fig:spreading_law}, which indicate robust decreasing trend of spreading rate with increasing liquid volatility ($\textcircled{\scriptsize{1}}$→$\textcircled{\scriptsize{3}}$→$\textcircled{\scriptsize{6}}$, $\textcircled{\scriptsize{2}}$→$\textcircled{\scriptsize{5}}$→$\textcircled{\scriptsize{8}}$), increasing substrate conductivity ($\textcircled{\scriptsize{3}}$→$\textcircled{\scriptsize{4}}$→$\textcircled{\scriptsize{5}}$, $\textcircled{\scriptsize{6}}$→$\textcircled{\scriptsize{7}}$→$\textcircled{\scriptsize{8}}$), and decreasing substrate thickness (from 3 mm copper to 0.3 mm copper). 

We further apply the calculated dimensionless numbers to the proposed model, and get the spreading component at the steady spreading stage. Very nice quantitative correspondence is found between experimental and theoretical values as summarized in Table \ref{tab:num_vs_exp}. Here, the experimental liquids and substrates cover most of the cases that span from non-volatile to most volatile liquids in practical applications and on substrates with a full range of thermal conductivity, thus allowing for reliable prediction of the liquid behaviors by quickly locating its Ja-Bi coordinate in the phase diagram. 

\begin{table}
\begin{center}
\def~{\hphantom{0}}
\begin{tabular}{ccccccccc}
 No. & Exp. & Num. & No. & Exp. & Num. & No. & Exp. & Num. \\
\hline
$\textcircled{\scriptsize{1}}$ & 0.0995±0.0028 & 0.103&$\textcircled{\scriptsize{2}}$ & 0.092±0.0028 & 0.0936&$\textcircled{\scriptsize{3}}$ & 0.0894±0.0026 & 0.0910 \\
$\textcircled{\scriptsize{4}}$ & 0.0869±0.0026 & 0.0813&$\textcircled{\scriptsize{5}}$ & 0.0803±0.0053 & 0.0788&$\textcircled{\scriptsize{6}}$ & 0.0634±0.0024 & 0.0752\\
$\textcircled{\scriptsize{7}}$ & 0.0601±0.0011 & 0.0578 &$\textcircled{\scriptsize{8}}$& 0.0568±0.0052 & 0.0504 & $\textcircled{\scriptsize{9}}$ & 0.0481±0.0036 & 0.0490\\
$\textcircled{\scriptsize{10}}$ & 0.0473±0.0030 & 0.0446&$\textcircled{\scriptsize{11}}$ & 0.0429±0.0010 & 0.0435 & 
\end{tabular}
\caption{\label{tab:num_vs_exp}
Experimental (Exp.) spreading exponent \emph{n} in comparison to the numerical (Num.) value by matching its position (Ja, Bi) in the phase diagram. Numbers $\textcircled{\scriptsize{1}}\sim\textcircled{\scriptsize{11}}$ correspond with the cases marked in \ref{fig:spreading_law}. Deviations between experimental and numerical results exist mainly in the high-volatility and low-substrate-conductivity region (small Ja and large Bi, \emph{e.g.}, case $\textcircled{\scriptsize{6}}$), where the heat dissipation into the gas phase is no longer negligible.}
\end{center}
\end{table}

Besides the experiments in present research, our conclusions on the effect of substrate thermal properties correspond with the work of \cite{shiri2021thermal} where alkane droplets indicate increasing contact angle and retarded spreading as the thickness of the glass substrate decreases stepwise from 6 mm to 0.063 mm. The influence of liquid volatility, we recently found, also meet with a very recent work by \cite{kumar2022wetting} (APS DFD abstract) where the spreading rate of HFE-7500 to 7100 droplets decreases with decreasing chain length, that is, increasing liquid volatility.

Looking at the overall life span of a deposited droplet, the experiments additionally indicate shortening spreading time of an evaporating droplet with increasing substrate conductivity, increasing liquid volatility, and increasing substrate temperature. For highly volatile liquids (lower left corner in the phase diagram), the spreading stage even disappears (indistinguishable) as the strong evaporation effect counteracts the capillary spreading, in which cases the contact line recedes right after the droplet deposition. This is exactly the case according to our observation of FC-72 droplets on heated substrates, where spreading slows down and the spreading period shortens as the substrate temperature increases stepwise from $25^\circ \rm C$ to $45^\circ \rm C$ as exampled by $\textcircled{\scriptsize{9}}$→$\textcircled{\scriptsize{10}}$→$\textcircled{\scriptsize{11}}$  in Fig. \ref{fig:spreading_law}. When the substrate temperature exceeds $45^\circ \rm C$ ($T_{\rm FC-72, boiling} = 56^\circ \rm C$ ), the spreading stage becomes indistinguishable, \emph{i.e.} the contact line starts to recede right after deposition, and the droplet ($\rm 0.6\ \mu L$) fully evaporates within 3 seconds, which represents a general scenario in flash evaporation.

\section{Conclusions and Perspectives}\label{sec:conclusions}
In this research, we revisit the classical problem of droplet wetting with interfacial phase change, which is of fundamental importance to a wide range of industrial processes that involve three phases. The objective originates from the urgent need for a unified explanation to a couple of basic scientific problems including the flow transition for different liquid-solid combinations, the existence of stagnation point near the three phase contact line, the relative strength of capillary and Marangoni flows, the deposition patterns from colloidal suspensions, as well as the corresponding spreading dynamics in diverse scenarios. Due to the requirement for substrate transparency and the available frame rate of existing visualization techniques, explorations on these scientific issues are limited to substrates like glass, and common liquids like water and alcohol, while important liquids and substrates in the vast industrial fields are less explored, such as refrigerants and metallic substrates. This requires a close collaboration between experimental and numerical efforts, where information that cannot be revealed from one side can be disclosed by the other.

Here, we develop a lubrication type model that considers the non-isothermal heat conduction into the solid substrate. We further conducted experiments on the flow field and the spreading rate for a wide range of combinations of liquids and solids. With one-to-one comparison between the numerical and experimental results, we elucidate the key dimensionless numbers that come into effect, \emph{i.e.}, the Jacok number, Ja, the liquid-solid Biot number, Bi, and the Marangoni number, Ma. A unified criteria is proposed for the transition of flow state, and provide unbiased explanations to a series of individual works on deposition patterns from drying droplets. A phase diagram for spreading dynamics is further established along with experimental demonstrations, which generalizes Tanner’s law (for non-volatile liquids in an isotherm state) to a full range of liquids with saturation vapor pressure spanning from $10^1$ to $10^4$ Pa and on substrates with all range of thermal conductivity spanning from $10^{-1}$ to $10^3$ W/m/K. 

In the theoretical aspect, this work illustrates the relative strength of interacting physical mechanisms as Regime I, II, III in Fig. \ref{fig:flow_reversal}, where the capillary effect dominates at low liquid volatility in Regime I, the Marangoni effect, capillary effect and evaporation effect interact in Regime II, and the strong evaporation effect (preferential mass loss near contact line) dominates in Regime III at very high liquid volatility. Such decomposition of the mechanisms allows for quick location of the influencing factors, and enables efficient control of liquid behaviors in various phase change energy devices, printing-coating processes, and microfluidic systems.

Overall, Tanner's law describes the spreading of non-volatile Newtonian fluids on smooth solid substrates. Here 'non-volatile', 'Newtonian' and 'smooth' are necessary conditions for the universality of the spreading exponent. With interfacial phase change (volatile liquids), the process becomes non-isothermal in which case the non-uniform distribution of mass flux and the thermal Marangoni effect come into play along with the capillary and viscous effect, as quantified in the present work. For non-Newtonian fluids, the viscosity largely depends on the shear stress. Theories for the spreading dynamics of dilatant fluids (shear thickening) and pseudo-plastic fluids (shear thinning) were established in previous studies based on the lubrication approximation (\cite{rafai2004spreading}, \cite{wang2007spreadingDynamics}) and experiments (\cite{wang2007spreading}). For topological surfaces with micro structures, the spreading dynamics can be more complex. A systematic understanding is yet reached due to the diverse surface topology and the complex liquid-solid interaction, where the small-scale surface structure affects the dynamics of the contact line and thereafter the overall droplet behaviors. By taking advantage of the asymmetric effect of surface topology, it is also possible to realize directional fluid motion, which indeed has been widely exploited in recent years with promising applications in directional liquid transport (\cite{li2017topological}, \cite{li2020bioinspired}), water harvesting (\cite{zheng2010directional}, \cite{dai2018hydrophilic}), and dry-out prevention (\cite{li2021directional}).

\vspace{20pt}
The authors gratefully acknowledge the support received from Japanese Society for the Promotion of Science (JSPS), and ThermaSMART project of European Commission (Grant No. EC-H2020-RISE-ThermaSMART-778104). ZW and CI acknowledge the support from JSPS KAKENHI (Grant Nos. JP21K14097, JP21H01251, JP22K18771). 

\vspace{20pt}
\textbf{Declaration of Interests} The authors report no conflict of interest.
 
\appendix
\section{Detailed Experimental Results}\label{appA}
\begin{figure}
    \centering
    \includegraphics[width=8cm]{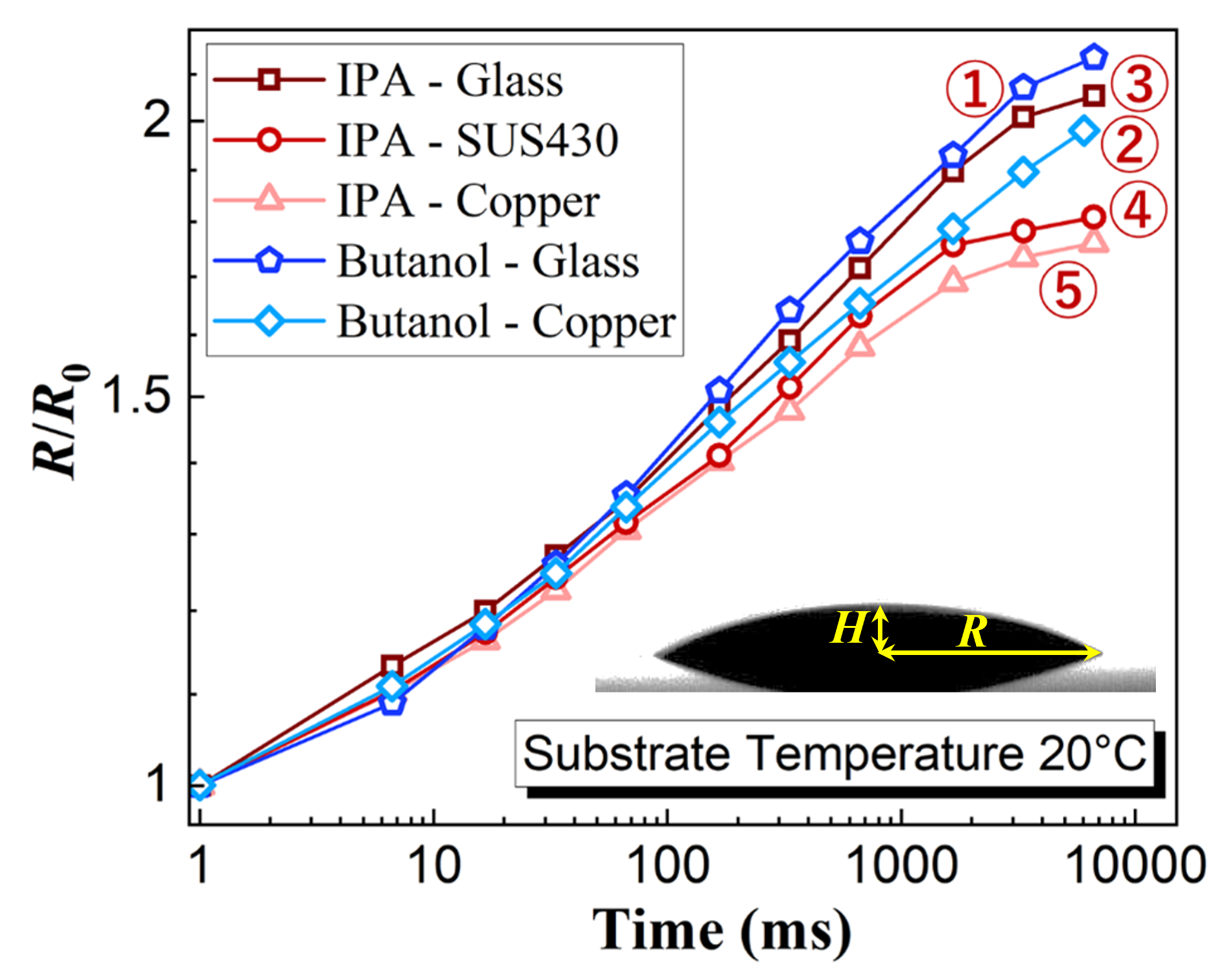}
    \caption{Spreading behaviors of droplets at substrates with different thermal conductivity ($T_\mathrm{sub}$ = 20°C). $\textcircled{\scriptsize{1}}$: IPA on 1.1 mm glass, $\textcircled{\scriptsize{2}}$: IPA on 1.0 mm SUS430, $\textcircled{\scriptsize{3}}$: IPA on 1.0 mm copper, $\textcircled{\scriptsize{4}}$: Butanol on 1.1 mm glass, $\textcircled{\scriptsize{5}}$: Butanol on 1 mm copper. These cases correspond with the mark numbers in Fig. \ref{fig:spreading_law}.}
    \label{fig:exp_results1}
\end{figure}

\begin{figure}
    \centering
    \includegraphics[width=8cm]{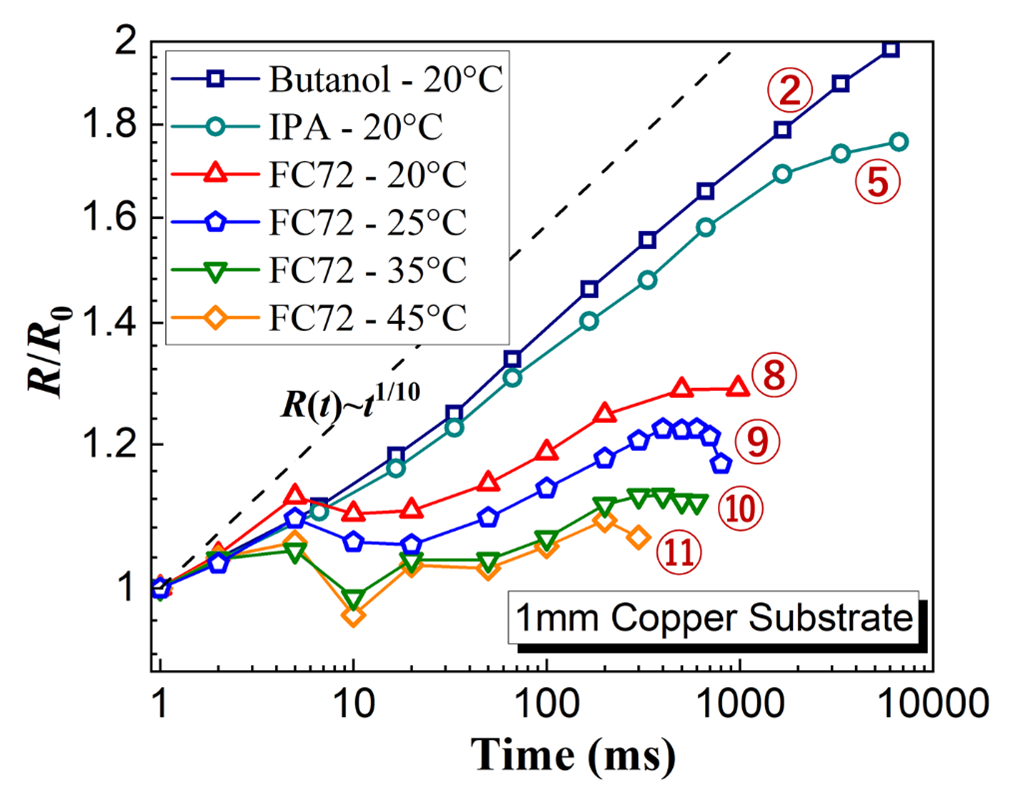}
    \caption{Spreading behaviors of droplets with different volatility on 1.0 mm copper substrate. $\textcircled{\scriptsize{2}}$: Butanol at $T_\mathrm{sub}$ = 20°C, $\textcircled{\scriptsize{5}}$: IPA at $T_\mathrm{sub}$ = 20°C, $\textcircled{\scriptsize{8}}\sim \textcircled{\scriptsize{10}}$: FC-72 at $T_\mathrm{sub}$ = 20°C, 25°C, 35°C, 45°C. These cases correspond with the mark numbers in Fig. \ref{fig:spreading_law}.}
    \label{fig:exp_results2}
\end{figure}

Fig. \ref{fig:exp_results1} and Fig. \ref{fig:exp_results2} present the variation of droplet contact radius along with time for representative combinations of liquids and solids (logarithmic coordinates). A summary and comparison of the spreading rate through experimental and numerical approaches are given in Table \ref{tab:num_vs_exp}. Taking an overview of all the variations curves, we can see that the effect of liquid volatility is much stronger than the substrate conductivity or thickness on the spreading dynamics of a droplet. 

For IPA and butanol droplets, spreading starts right after the droplet gets contact with the solid surface and lasts for $10^0\sim 10^1$ seconds before the contact line slows down and recedes. By changing the substrate material from glass to SUS430 and to copper as in Fig. \ref{fig:exp_results1}, the slope of the variation curve decreases, indicating a decreasing trend of spreading rate with increasing substrate conductivity ($\textcircled{\scriptsize{1}}$→$\textcircled{\scriptsize{2}}$, $\textcircled{\scriptsize{3}}$→$\textcircled{\scriptsize{4}}$→$\textcircled{\scriptsize{5}}$).

Fig. \ref{fig:exp_results2} indicates the effect of liquid volatility on droplet spreading. For FC-72 droplets, a slight fluctuation (spread and contract) takes place at the very initial moment ($< 10\ ms$) when a droplet contacts the solid surface as a result of the inertia effect (even through the distance between the syringe tip and the solid surface is controlled as small as possible, the inertia effect can cause a slight fluctuation when the liquid surface tension is small, which can be evaluated by the Weber number). The droplet ($\textcircled{\scriptsize{8}}$ in Fig. \ref{fig:exp_results2}) then spreads, slows down and recedes after several hundred milliseconds. By further increasing the substrate temperature, the period of spreading shortens along with decreasing spreading rate as indicated by $\textcircled{\scriptsize{8}}$→$\textcircled{\scriptsize{9}}$→$\textcircled{\scriptsize{10}}$→$\textcircled{\scriptsize{11}}$ in Fig. \ref{fig:exp_results2}. For substrate temperature higher than 45 °C, the spreading stage is no longer distinguishable, in which case the evaporation effect is strong enough to counteract the capillary spreading. As a result, no apparent spreading can be observed, \emph{i.e.}, the contact line recedes right after the droplet deposition due to significant mass loss at the contact line region.

In the mathematical model, we formulate the expression of interfacial mass flux by combining the Hertz-Knudsen equation, the chemical balance relation across the liquid-air interface, and the ideal gas assumption. The Hertz-Knudsen equation is often utilized to describe the evaporation rates based on the statistics of vapor molecules adsorbed onto the liquid surface. The evaporation and condensation coefficients contained in the equation are thus important while they have inexplicably been found to span 3 orders of magnitude when comparing with the actual experimental data as reported in a number of studies (\cite{persad2016expressions}, \cite{young1991condensation}, \cite{fang1999temperature}). In the formulation, we set the evaporation and condensation coefficients as 1 with the assumption that the phase change process is near equilibrium. A modified Ja number, Ja*, is utilized for numerical calculations to compensate for the overprediction of the Hertz-Knudsen equation, where liquids with higher volatility (higher molecular activity) tend to be more overpredicted. Specifically, Ja* is calculated by magnifying Ja for $500\frac{\mathrm{log}_\mathrm{max}-\mathrm{log}Ja}{\mathrm{log}_\mathrm{max}-\mathrm{log}_\mathrm{min}}$ times, where $\mathrm{log}_\mathrm{max}$ is the maximal order of magnitude of Ja for all possible types of liquids and $\mathrm{log}_\mathrm{min}$ is the minimal order of magnitude ($\mathrm{log}_\mathrm{max}=0$, $\mathrm{log}_\mathrm{min}=-4$ in our present work). The calculated values of Ja* and Bi for cases $\textcircled{\scriptsize{1}}$ to $\textcircled{\scriptsize{11}}$ are summarized in Table \ref{tab:value_Ja_Bi}, which correspond with the data points marked in phase diagram Fig. \ref{fig:spreading_law}.

\begin{table}
  \begin{center}
\def~{\hphantom{0}}
  \begin{tabular}{cccccccccccc}
Case& $\textcircled{\scriptsize{1}}$&$\textcircled{\scriptsize{2}}$ &$\textcircled{\scriptsize{3}}$ &$\textcircled{\scriptsize{4}}$ &$\textcircled{\scriptsize{5}}$ &$\textcircled{\scriptsize{6}}$ &$\textcircled{\scriptsize{7}}$ &$\textcircled{\scriptsize{8}}$ &$\textcircled{\scriptsize{9}}$ &$\textcircled{\scriptsize{10}}$ &$\textcircled{\scriptsize{11}}$\\
Bi&2.8e-1&	8.1e-4&	2.5e-1&	1.1e-2&	7.0e-4&	1.7e-1&	7.6e-3&	4.9e-4&	4.9e-4&	4.9e-4&	4.9e-4\\
Ja*& 7.0e-1&	7.0e-1&	1.1e-1&	1.1e-1&	1.1e-1&	1.9e-2&	1.9e-2&	1.9e-2&	1.6e-2&	1.1e-2&	7.9e-3
  \end{tabular}
\caption{Values of Ja* and Bi for cases $\textcircled{\scriptsize{1}}$ to $\textcircled{\scriptsize{11}}$ (expressed in scientific notation).}
  \label{tab:value_Ja_Bi}
  \end{center}
\end{table}

Additional PIV experiments indicate the transition of flow field as the liquid volatility increases. For low-volatility Butanol, continuous outward flow is observed driven by the capillary force, which forms a ‘coffee ring’ as the droplet dries out (Fig. \ref{fig:microPIV}(a), Supplementary Movie 3, left). As the liquid volatility increases (IPA), the PIV particles tend to change its direction at a position near the three-phase contact line (stagnation point) as shown by the tracing track of fluorescent particles in Fig. \ref{fig:microPIV}(b). The direction change avoids particle accumulation at the periphery, which leads to more uniform deposition patterns as the droplet fully dries out. The transition of flow direction and the formation of deposition patterns correspond with our numerical predictions by locating the Ja-Bi positions for (1) Butanol-slide glass and (2) IPA-slide glass in Fig. \ref{fig:flow_reversal}.

\bibliographystyle{jfm}
\bibliography{jfm-instructions}

\providecommand{\noopsort}[1]{}\providecommand{\singleletter}[1]{#1}%
\begin{thebibliography}{62}
\expandafter\ifx\csname natexlab\endcsname\relax\def\natexlab#1{#1}\fi
\def\au#1{#1} \def\ed#1{#1} \def\yr#1{#1}\def\at#1{#1}\def\jt#1{\textit{#1}}
  \def\bt#1{#1}\def\bvol#1{\textbf{#1}} \def\vol#1{#1} \def\pg#1{#1}
  \def\publ#1{#1}\def\arxiv#1{#1}\def\org#1{#1}\def\st#1{\textit{#1}}

\bibitem[Ajaev(2005)]{ajaev2005spreading}
{\sc \au{Ajaev, Vladimir~S}} \yr{2005}  \at{Spreading of thin volatile liquid
  droplets on uniformly heated surfaces}.  \jt{Journal of Fluid Mechanics}
  \bvol{528},  \pg{279--296}.

\bibitem[Anderson \& Davis(1995)]{anderson1995spreading}
{\sc \au{Anderson, DM} \& \au{Davis, SH}} \yr{1995}  \at{The spreading of
  volatile liquid droplets on heated surfaces}.  \jt{Physics of Fluids}
  \bvol{7}~(2),  \pg{248--265}.

\bibitem[Atkins {\em et~al.\/}(2014)Atkins, Atkins \&
  de~Paula]{atkins2014atkins}
{\sc \au{Atkins, Peter}, \au{Atkins, Peter~William} \& \au{de~Paula, Julio}}
  \yr{2014} {\em Atkins' physical chemistry\/}.  \publ{Oxford university
  press}.

\bibitem[Berteloot {\em et~al.\/}(2008)Berteloot, Pham, Daerr, Lequeux \&
  Limat]{berteloot2008evaporation}
{\sc \au{Berteloot, Guillaume}, \au{Pham, C-T}, \au{Daerr, A}, \au{Lequeux,
  Fran{\c{c}}ois} \& \au{Limat, Laurent}} \yr{2008}  \at{Evaporation-induced
  flow near a contact line: Consequences on coating and contact angle}.
  \jt{Europhysics Letters}  \bvol{83}~(1),  \pg{14003}.

\bibitem[Blake(2006)]{blake2006physics}
{\sc \au{Blake, Terence~D}} \yr{2006}  \at{The physics of moving wetting
  lines}.  \jt{Journal of Colloid and Interface Science}  \bvol{299}~(1),
  \pg{1--13}.

\bibitem[Bonn {\em et~al.\/}(2009)Bonn, Eggers, Indekeu, Meunier \&
  Rolley]{bonn2009wetting}
{\sc \au{Bonn, Daniel}, \au{Eggers, Jens}, \au{Indekeu, Joseph}, \au{Meunier,
  Jacques} \& \au{Rolley, Etienne}} \yr{2009}  \at{Wetting and spreading}.
  \jt{Reviews of Modern Physics}  \bvol{81}~(2),  \pg{739}.

\bibitem[Breitenbach {\em et~al.\/}(2018)Breitenbach, Roisman \&
  Tropea]{breitenbach2018drop}
{\sc \au{Breitenbach, Jan}, \au{Roisman, Ilia~V} \& \au{Tropea, Cameron}}
  \yr{2018}  \at{From drop impact physics to spray cooling models: a critical
  review}.  \jt{Experiments in Fluids}  \bvol{59},  \pg{1--21}.

\bibitem[Carlson(2018)]{carlson2018fluctuation}
{\sc \au{Carlson, Andreas}} \yr{2018}  \at{Fluctuation assisted spreading of a
  fluid filled elastic blister}.  \jt{Journal of Fluid Mechanics}  \bvol{846},
  \pg{1076--1087}.

\bibitem[Carr{\'e} {\em et~al.\/}(1996)Carr{\'e}, Gastel \&
  Shanahan]{carre1996viscoelastic}
{\sc \au{Carr{\'e}, Alain}, \au{Gastel, Jean-Claude} \& \au{Shanahan,
  Martin~ER}} \yr{1996}  \at{Viscoelastic effects in the spreading of liquids}.
   \jt{Nature}  \bvol{379}~(6564),  \pg{432--434}.

\bibitem[Cazabat \& Stuart(1986)]{cazabat1986dynamics}
{\sc \au{Cazabat, AM} \& \au{Stuart, MA~Cohen}} \yr{1986}  \at{Dynamics of
  wetting: effects of surface roughness}.  \jt{The Journal of Physical
  Chemistry}  \bvol{90}~(22),  \pg{5845--5849}.

\bibitem[Charitatos \& Kumar(2020)]{charitatos2020thin}
{\sc \au{Charitatos, Vasileios} \& \au{Kumar, Satish}} \yr{2020}  \at{A
  thin-film model for droplet spreading on soft solid substrates}.  \jt{Soft
  Matter}  \bvol{16}~(35),  \pg{8284--8298}.

\bibitem[Cormier {\em et~al.\/}(2012)Cormier, McGraw, Salez, Rapha{\"e}l \&
  Dalnoki-Veress]{cormier2012beyond}
{\sc \au{Cormier, Sara~L}, \au{McGraw, Joshua~D}, \au{Salez, Thomas},
  \au{Rapha{\"e}l, Elie} \& \au{Dalnoki-Veress, Kari}} \yr{2012}  \at{Beyond
  tanner’s law: Crossover between spreading regimes of a viscous droplet on
  an identical film}.  \jt{Physical Review Letters}  \bvol{109}~(15),
  \pg{154501}.

\bibitem[Dai {\em et~al.\/}(2016)Dai, Khonsari, Shen, Huang \&
  Wang]{dai2016thermocapillary}
{\sc \au{Dai, Qingwen}, \au{Khonsari, MM}, \au{Shen, Cong}, \au{Huang, Wei} \&
  \au{Wang, Xiaolei}} \yr{2016}  \at{Thermocapillary migration of liquid
  droplets induced by a unidirectional thermal gradient}.  \jt{Langmuir}
  \bvol{32}~(30),  \pg{7485--7492}.

\bibitem[Dai {\em et~al.\/}(2018)Dai, Sun, Nielsen, Stogin, Wang, Yang \&
  Wong]{dai2018hydrophilic}
{\sc \au{Dai, Xianming}, \au{Sun, Nan}, \au{Nielsen, Steven~O}, \au{Stogin,
  Birgitt~Boschitsch}, \au{Wang, Jing}, \au{Yang, Shikuan} \& \au{Wong,
  Tak-Sing}} \yr{2018}  \at{Hydrophilic directional slippery rough surfaces for
  water harvesting}.  \jt{Science advances}  \bvol{4}~(3),  \pg{eaaq0919}.

\bibitem[De~Gennes {\em et~al.\/}(1990)De~Gennes, Hua \&
  Levinson]{de1990dynamics}
{\sc \au{De~Gennes, PG}, \au{Hua, X} \& \au{Levinson, P}} \yr{1990}
  \at{Dynamics of wetting: local contact angles}.  \jt{Journal of fluid
  mechanics}  \bvol{212},  \pg{55--63}.

\bibitem[De~Gennes(1985)]{de1985wetting}
{\sc \au{De~Gennes, Pierre-Gilles}} \yr{1985}  \at{Wetting: statics and
  dynamics}.  \jt{Reviews of Modern Physics}  \bvol{57}~(3),  \pg{827}.

\bibitem[Deegan {\em et~al.\/}(1997)Deegan, Bakajin, Dupont, Huber, Nagel \&
  Witten]{deegan1997capillary}
{\sc \au{Deegan, Robert~D}, \au{Bakajin, Olgica}, \au{Dupont, Todd~F},
  \au{Huber, Greb}, \au{Nagel, Sidney~R} \& \au{Witten, Thomas~A}} \yr{1997}
  \at{Capillary flow as the cause of ring stains from dried liquid drops}.
  \jt{Nature}  \bvol{389}~(6653),  \pg{827--829}.

\bibitem[Dunn {\em et~al.\/}(2009)Dunn, Wilson, Duffy, David \&
  Sefiane]{dunn2009strong}
{\sc \au{Dunn, GJ}, \au{Wilson, SK}, \au{Duffy, BR}, \au{David, S} \&
  \au{Sefiane, Khellil}} \yr{2009}  \at{The strong influence of substrate
  conductivity on droplet evaporation}.  \jt{Journal of Fluid Mechanics}
  \bvol{623},  \pg{329--351}.

\bibitem[Fang \& Ward(1999)]{fang1999temperature}
{\sc \au{Fang, G} \& \au{Ward, CA}} \yr{1999}  \at{Temperature measured close
  to the interface of an evaporating liquid}.  \jt{Physical Review E}
  \bvol{59}~(1),  \pg{417}.

\bibitem[Goossens {\em et~al.\/}(2011)Goossens, Seveno, Rioboo, Vaillant, Conti
  \& De~Coninck]{goossens2011can}
{\sc \au{Goossens, Sarah}, \au{Seveno, David}, \au{Rioboo, Romain},
  \au{Vaillant, Alexandre}, \au{Conti, J} \& \au{De~Coninck, Jo{\"e}l}}
  \yr{2011}  \at{Can we predict the spreading of a two-liquid system from the
  spreading of the corresponding liquid--air systems?}  \jt{Langmuir}
  \bvol{27}~(16),  \pg{9866--9872}.

\bibitem[Hoang \& Kavehpour(2011)]{hoang2011dynamics}
{\sc \au{Hoang, A} \& \au{Kavehpour, HP}} \yr{2011}  \at{Dynamics of nanoscale
  precursor film near a moving contact line of spreading drops}.  \jt{Physical
  review letters}  \bvol{106}~(25),  \pg{254501}.

\bibitem[Hu \& Larson(2005)]{hu2005analysis}
{\sc \au{Hu, Hua} \& \au{Larson, Ronald~G}} \yr{2005}  \at{Analysis of the
  effects of marangoni stresses on the microflow in an evaporating sessile
  droplet}.  \jt{Langmuir}  \bvol{21}~(9),  \pg{3972--3980}.

\bibitem[Hu \& Larson(2006)]{hu2006marangoni}
{\sc \au{Hu, Hua} \& \au{Larson, Ronald~G}} \yr{2006}  \at{Marangoni effect
  reverses coffee-ring depositions}.  \jt{The Journal of Physical Chemistry B}
  \bvol{110}~(14),  \pg{7090--7094}.

\bibitem[Jalaal {\em et~al.\/}(2021)Jalaal, Stoeber \&
  Balmforth]{jalaal2021spreading}
{\sc \au{Jalaal, Maziyar}, \au{Stoeber, Boris} \& \au{Balmforth, Neil~J}}
  \yr{2021}  \at{Spreading of viscoplastic droplets}.  \jt{Journal of Fluid
  Mechanics}  \bvol{914},  \pg{A21}.

\bibitem[Jambon-Puillet {\em et~al.\/}(2018)Jambon-Puillet, Carrier,
  Shahidzadeh, Brutin, Eggers \& Bonn]{jambon2018spreading}
{\sc \au{Jambon-Puillet, Etienne}, \au{Carrier, Odile}, \au{Shahidzadeh,
  Noushine}, \au{Brutin, David}, \au{Eggers, Jens} \& \au{Bonn, Daniel}}
  \yr{2018}  \at{Spreading dynamics and contact angle of completely wetting
  volatile drops}.  \jt{Journal of Fluid Mechanics}  \bvol{844},
  \pg{817--830}.

\bibitem[Karapetsas {\em et~al.\/}(2013)Karapetsas, Sahu \&
  Matar]{karapetsas2013effect}
{\sc \au{Karapetsas, George}, \au{Sahu, Kirti~Chandra} \& \au{Matar, Omar~K}}
  \yr{2013}  \at{Effect of contact line dynamics on the thermocapillary motion
  of a droplet on an inclined plate}.  \jt{Langmuir}  \bvol{29}~(28),
  \pg{8892--8906}.

\bibitem[Kavehpour {\em et~al.\/}(2003)Kavehpour, Ovryn \&
  McKinley]{kavehpour2003microscopic}
{\sc \au{Kavehpour, H~Pirouz}, \au{Ovryn, Ben} \& \au{McKinley, Gareth~H}}
  \yr{2003}  \at{Microscopic and macroscopic structure of the precursor layer
  in spreading viscous drops}.  \jt{Physical review letters}  \bvol{91}~(19),
  \pg{196104}.

\bibitem[Kumar {\em et~al.\/}(2022)Kumar, Parimalanathan, Rednikov \&
  Colinet]{kumar2022wetting}
{\sc \au{Kumar, Manish}, \au{Parimalanathan, Senthil}, \au{Rednikov, Alexey} \&
  \au{Colinet, Pierre}} \yr{2022}  \at{Wetting dynamics of volatile
  hydrofluoroether (hfe) liquid droplets}.  \jt{Bulletin of the American
  Physical Society} .

\bibitem[Lelah \& Marmur(1981)]{lelah1981spreading}
{\sc \au{Lelah, Michael~D} \& \au{Marmur, Abraham}} \yr{1981}  \at{Spreading
  kinetics of drops on glass}.  \jt{Journal of Colloid and Interface Science}
  \bvol{82}~(2),  \pg{518--525}.

\bibitem[Li {\em et~al.\/}(2017)Li, Zhou, Li, Che, Yao, McHale, Chaudhury \&
  Wang]{li2017topological}
{\sc \au{Li, Jiaqian}, \au{Zhou, Xiaofeng}, \au{Li, Jing}, \au{Che, Lufeng},
  \au{Yao, Jun}, \au{McHale, Glen}, \au{Chaudhury, Manoj~K} \& \au{Wang,
  Zuankai}} \yr{2017}  \at{Topological liquid diode}.  \jt{Science advances}
  \bvol{3}~(10),  \pg{eaao3530}.

\bibitem[Li {\em et~al.\/}(2021)Li, Zhou, Tao, Zheng \&
  Wang]{li2021directional}
{\sc \au{Li, Jiaqian}, \au{Zhou, Xiaofeng}, \au{Tao, Ran}, \au{Zheng, Huanxi}
  \& \au{Wang, Zuankai}} \yr{2021}  \at{Directional liquid transport from the
  cold region to the hot region on a topological surface}.  \jt{Langmuir}
  \bvol{37}~(16),  \pg{5059--5065}.

\bibitem[Li {\em et~al.\/}(2020)Li, Li \& Dong]{li2020bioinspired}
{\sc \au{Li, Xing}, \au{Li, Jiaqian} \& \au{Dong, Guangneng}} \yr{2020}
  \at{Bioinspired topological surface for directional oil lubrication}.
  \jt{ACS applied materials \& interfaces}  \bvol{12}~(4),  \pg{5113--5119}.

\bibitem[Li {\em et~al.\/}(2015)Li, Lv, Li, Qu{\'e}r{\'e} \&
  Zheng]{li2015coffee}
{\sc \au{Li, Yanshen}, \au{Lv, Cunjing}, \au{Li, Zhaohan}, \au{Qu{\'e}r{\'e},
  David} \& \au{Zheng, Quanshui}} \yr{2015}  \at{From coffee rings to coffee
  eyes}.  \jt{Soft Matter}  \bvol{11}~(23),  \pg{4669--4673}.

\bibitem[Marmur(1983)]{marmur1983equilibrium}
{\sc \au{Marmur, Abraham}} \yr{1983}  \at{Equilibrium and spreading of liquids
  on solid surfaces}.  \jt{Advances in Colloid and Interface Science}
  \bvol{19}~(1-2),  \pg{75--102}.

\bibitem[Mitra \& Mitra(2016)]{mitra2016understanding}
{\sc \au{Mitra, Surjyasish} \& \au{Mitra, Sushanta~K}} \yr{2016}
  \at{Understanding the early regime of drop spreading}.  \jt{Langmuir}
  \bvol{32}~(35),  \pg{8843--8848}.

\bibitem[Moosman \& Homsy(1980)]{moosman1980evaporating}
{\sc \au{Moosman, Steven} \& \au{Homsy, GM}} \yr{1980}  \at{Evaporating menisci
  of wetting fluids}.  \jt{Journal of Colloid and Interface Science}
  \bvol{73}~(1),  \pg{212--223}.

\bibitem[Mudawar(2001)]{mudawar2001assessment}
{\sc \au{Mudawar, Issam}} \yr{2001}  \at{Assessment of high-heat-flux thermal
  management schemes}.  \jt{IEEE Transactions on Components and Packaging
  Technologies}  \bvol{24}~(2),  \pg{122--141}.

\bibitem[Pang \& {\'O}~N{\'a}raigh(2022)]{pang2022mathematical}
{\sc \au{Pang, Khang~Ee} \& \au{{\'O}~N{\'a}raigh, Lennon}} \yr{2022}  \at{A
  mathematical model and mesh-free numerical method for contact-line motion in
  lubrication theory}.  \jt{Environmental Fluid Mechanics}  \bvol{22}~(2-3),
  \pg{301--336}.

\bibitem[Persad \& Ward(2016)]{persad2016expressions}
{\sc \au{Persad, Aaron~H} \& \au{Ward, Charles~A}} \yr{2016}  \at{Expressions
  for the evaporation and condensation coefficients in the hertz-knudsen
  relation}.  \jt{Chemical reviews}  \bvol{116}~(14),  \pg{7727--7767}.

\bibitem[Plesset \& Prosperetti(1976)]{plesset1976flow}
{\sc \au{Plesset, Milton~S} \& \au{Prosperetti, Andrea}} \yr{1976}  \at{Flow of
  vapour in a liquid enclosure}.  \jt{Journal of Fluid Mechanics}
  \bvol{78}~(3),  \pg{433--444}.

\bibitem[Rafa{\"\i} {\em et~al.\/}(2004)Rafa{\"\i}, Bonn \&
  Boudaoud]{rafai2004spreading}
{\sc \au{Rafa{\"\i}, Salima}, \au{Bonn, Daniel} \& \au{Boudaoud, Arezki}}
  \yr{2004}  \at{Spreading of non-newtonian fluids on hydrophilic surfaces}.
  \jt{Journal of Fluid Mechanics}  \bvol{513},  \pg{77--85}.

\bibitem[Ristenpart {\em et~al.\/}(2007)Ristenpart, Kim, Domingues, Wan \&
  Stone]{ristenpart2007influence}
{\sc \au{Ristenpart, WD}, \au{Kim, PG}, \au{Domingues, C}, \au{Wan, J} \&
  \au{Stone, Howard~A}} \yr{2007}  \at{Influence of substrate conductivity on
  circulation reversal in evaporating drops}.  \jt{Physical Review Letters}
  \bvol{99}~(23),  \pg{234502}.

\bibitem[Rose(2002)]{rose2002dropwise}
{\sc \au{Rose, JW}} \yr{2002}  \at{Dropwise condensation theory and experiment:
  a review}.  \jt{Proceedings of the Institution of Mechanical Engineers, Part
  A: Journal of Power and Energy}  \bvol{216}~(2),  \pg{115--128}.

\bibitem[Saiseau {\em et~al.\/}(2022)Saiseau, Pedersen, Benjana, Carlson,
  Delabre, Salez \& Delville]{saiseau2022near}
{\sc \au{Saiseau, Raphael}, \au{Pedersen, Christian}, \au{Benjana, Anwar},
  \au{Carlson, Andreas}, \au{Delabre, Ulysse}, \au{Salez, Thomas} \&
  \au{Delville, Jean-Pierre}} \yr{2022}  \at{Near-critical spreading of
  droplets}.  \jt{Nature Communications}  \bvol{13}~(1),  \pg{7442}.

\bibitem[Schneider {\em et~al.\/}(2012)Schneider, Rasband \&
  Eliceiri]{schneider2012nih}
{\sc \au{Schneider, Caroline~A}, \au{Rasband, Wayne~S} \& \au{Eliceiri,
  Kevin~W}} \yr{2012}  \at{Nih image to imagej: 25 years of image analysis}.
  \jt{Nature methods}  \bvol{9}~(7),  \pg{671--675}.

\bibitem[Shiri {\em et~al.\/}(2021)Shiri, Sinha, Baumgartner \&
  Cira]{shiri2021thermal}
{\sc \au{Shiri, Samira}, \au{Sinha, Shayandev}, \au{Baumgartner, Dieter~A} \&
  \au{Cira, Nate~J}} \yr{2021}  \at{Thermal marangoni flow impacts the shape of
  single component volatile droplets on thin, completely wetting substrates}.
  \jt{Physical Review Letters}  \bvol{127}~(2),  \pg{024502}.

\bibitem[Tanner(1979)]{tanner1979spreading}
{\sc \au{Tanner, LH}} \yr{1979}  \at{The spreading of silicone oil drops on
  horizontal surfaces}.  \jt{Journal of Physics D: Applied Physics}
  \bvol{12}~(9),  \pg{1473}.

\bibitem[Teh {\em et~al.\/}(2008)Teh, Lin, Hung \& Lee]{teh2008droplet}
{\sc \au{Teh, Shia-Yen}, \au{Lin, Robert}, \au{Hung, Lung-Hsin} \& \au{Lee,
  Abraham~P}} \yr{2008}  \at{Droplet microfluidics}.  \jt{Lab on a Chip}
  \bvol{8}~(2),  \pg{198--220}.

\bibitem[Tsoumpas {\em et~al.\/}(2015)Tsoumpas, Dehaeck, Rednikov \&
  Colinet]{tsoumpas2015effect}
{\sc \au{Tsoumpas, Yannis}, \au{Dehaeck, Sam}, \au{Rednikov, Alexey} \&
  \au{Colinet, Pierre}} \yr{2015}  \at{Effect of marangoni flows on the shape
  of thin sessile droplets evaporating into air}.  \jt{Langmuir}
  \bvol{31}~(49),  \pg{13334--13340}.

\bibitem[Tu {\em et~al.\/}(2018)Tu, Wang, Zhang \& Wang]{tu2018progress}
{\sc \au{Tu, Yaodong}, \au{Wang, Ruzhu}, \au{Zhang, Yannan} \& \au{Wang,
  Jiayun}} \yr{2018}  \at{Progress and expectation of atmospheric water
  harvesting}.  \jt{Joule}  \bvol{2}~(8),  \pg{1452--1475}.

\bibitem[Voinov(1976)]{voinov1976hydrodynamics}
{\sc \au{Voinov, OV}} \yr{1976}  \at{Hydrodynamics of wetting}.  \jt{Fluid
  Dynamics}  \bvol{11}~(5),  \pg{714--721}.

\bibitem[Wang \& Harris(2018)]{wang2018stagnation}
{\sc \au{Wang, Lihui} \& \au{Harris, Michael~T}} \yr{2018}  \at{Stagnation
  point of surface flow during drop evaporation}.  \jt{Langmuir}
  \bvol{34}~(20),  \pg{5918--5925}.

\bibitem[Wang {\em et~al.\/}(2007{\natexlab{{\em a\/}}})Wang, Lee, Peng \&
  Lai]{wang2007spreadingDynamics}
{\sc \au{Wang, XD}, \au{Lee, DJ}, \au{Peng, XF} \& \au{Lai, JY}}
  \yr{2007{\natexlab{{\em a\/}}}}  \at{Spreading dynamics and dynamic contact
  angle of non-newtonian fluids}.  \jt{Langmuir}  \bvol{23}~(15),
  \pg{8042--8047}.

\bibitem[Wang {\em et~al.\/}(2007{\natexlab{{\em b\/}}})Wang, Zhang, Lee \&
  Peng]{wang2007spreading}
{\sc \au{Wang, XD}, \au{Zhang, Y}, \au{Lee, DJ} \& \au{Peng, XF}}
  \yr{2007{\natexlab{{\em b\/}}}}  \at{Spreading of completely wetting or
  partially wetting power-law fluid on solid surface}.  \jt{Langmuir}
  \bvol{23}~(18),  \pg{9258--9262}.

\bibitem[Wang {\em et~al.\/}(2021)Wang, Karapetsas, Valluri, Sefiane, Williams
  \& Takata]{wang2021dynamics}
{\sc \au{Wang, Zhenying}, \au{Karapetsas, George}, \au{Valluri, Prashant},
  \au{Sefiane, Khellil}, \au{Williams, Adam} \& \au{Takata, Yasuyuki}}
  \yr{2021}  \at{Dynamics of hygroscopic aqueous solution droplets undergoing
  evaporation or vapour absorption}.  \jt{Journal of Fluid Mechanics}
  \bvol{912},  \pg{A2}.

\bibitem[Wang {\em et~al.\/}(2019)Wang, Orejon, Sefiane \&
  Takata]{wang2019coupled}
{\sc \au{Wang, Zhenying}, \au{Orejon, Daniel}, \au{Sefiane, Khellil} \&
  \au{Takata, Yasuyuki}} \yr{2019}  \at{Coupled thermal transport and mass
  diffusion during vapor absorption into hygroscopic liquid desiccant
  droplets}.  \jt{International Journal of Heat and Mass Transfer}  \bvol{134},
   \pg{1014--1023}.

\bibitem[Wang {\em et~al.\/}(2022)Wang, Orejon, Takata \&
  Sefiane]{wang2022wetting}
{\sc \au{Wang, Zhenying}, \au{Orejon, Daniel}, \au{Takata, Yasuyuki} \&
  \au{Sefiane, Khellil}} \yr{2022}  \at{Wetting and evaporation of
  multicomponent droplets}.  \jt{Physics Reports}  \bvol{960},  \pg{1--37}.

\bibitem[Xu \& Luo(2007)]{xu2007marangoni}
{\sc \au{Xu, Xuefeng} \& \au{Luo, Jianbin}} \yr{2007}  \at{Marangoni flow in an
  evaporating water droplet}.  \jt{Applied Physics Letters}  \bvol{91}~(12),
  \pg{124102}.

\bibitem[Xu {\em et~al.\/}(2010)Xu, Luo \& Guo]{xu2010criterion}
{\sc \au{Xu, Xuefeng}, \au{Luo, Jianbin} \& \au{Guo, Dan}} \yr{2010}
  \at{Criterion for reversal of thermal marangoni flow in drying drops}.
  \jt{Langmuir}  \bvol{26}~(3),  \pg{1918--1922}.

\bibitem[Yang {\em et~al.\/}(2022)Yang, Wu, Doi \& Man]{yang2022evaporation}
{\sc \au{Yang, Xiuyuan}, \au{Wu, MengMeng}, \au{Doi, Masao} \& \au{Man,
  Xingkun}} \yr{2022}  \at{Evaporation dynamics of sessile droplets: The
  intricate coupling of capillary, evaporation, and marangoni flow}.
  \jt{Langmuir}  \bvol{38}~(16),  \pg{4887--4893}.

\bibitem[Young(1991)]{young1991condensation}
{\sc \au{Young, JB}} \yr{1991}  \at{The condensation and evaporation of liquid
  droplets in a pure vapour at arbitrary knudsen number}.  \jt{International
  journal of heat and mass transfer}  \bvol{34}~(7),  \pg{1649--1661}.

\bibitem[Zheng {\em et~al.\/}(2010)Zheng, Bai, Huang, Tian, Nie, Zhao, Zhai \&
  Jiang]{zheng2010directional}
{\sc \au{Zheng, Yongmei}, \au{Bai, Hao}, \au{Huang, Zhongbing}, \au{Tian,
  Xuelin}, \au{Nie, Fu-Qiang}, \au{Zhao, Yong}, \au{Zhai, Jin} \& \au{Jiang,
  Lei}} \yr{2010}  \at{Directional water collection on wetted spider silk}.
  \jt{Nature}  \bvol{463}~(7281),  \pg{640--643}.

\end{thebibliography}

\end{document}